\documentclass[12pt]{article}
\usepackage{amsmath, amssymb, amsfonts, amsthm}
\usepackage{graphicx}
\usepackage{enumerate}
\usepackage{natbib}

\usepackage{url} 
\numberwithin{equation}{section}

\newcommand{\blind}{1}

\newtheoremstyle{break}
  {\topsep}{\topsep}%
  {\itshape}{}%
  {\bfseries}{}%
  {\newline}{}%
\theoremstyle{break}
\newtheorem{theorem}{Theorem}

\newtheorem{assumption}{Assumption}

\newcommand{\bDelta}{ \mbox{\boldmath $\Delta$} }

\newcommand{\blambda}{ \mbox{\boldmath $\lambda$} }

\newcommand{\bbeta}{ \mbox{\boldmath $\beta$} }

\newcommand{\bg}{ {\bf g} }

\newcommand{\bX}{ {\bf X} }

\newcommand{\bW}{ {\bf W} }

\newcommand{\bV}{ {\bf V} }

\newcommand{\bE}{ {\bf E} }

\newcommand{\ind}{\perp\!\!\!\!\perp}

\begin{document}

\def\spacingset#1{\renewcommand{\baselinestretch}%
{#1}\small\normalsize} \spacingset{1}


\if1\blind
{
  \title{\bf Inference of Treatment Effect and Its Regional Modifiers Using Restricted Mean Survival Time in Multi-Regional Clinical Trials}
  \author{Kaiyuan Hua$^a$, Hwanhee Hong$^a$, and Xiaofei Wang$^a$\\
    $^a$Department of Biostatistics and Bioinformatics, \\
    Duke University School of Medicine}
  \maketitle
} \fi

\if0\blind
{
  \bigskip
  \bigskip
  \bigskip
  \begin{center}
    {\LARGE\bf Inference of Treatment Effect and Its Regional Modifiers Using Restricted Mean Survival Time in Multi-Regional Clinical Trials}
\end{center}
  \medskip
} \fi

\bigskip
\begin{abstract}
Multi-regional clinical trials (MRCTs) play an increasingly crucial role in global pharmaceutical development by expediting data gathering and regulatory approval across diverse patient populations. However, differences in recruitment practices and regional demographics often lead to variations in study participant characteristics, potentially biasing treatment effect estimates and undermining treatment effect consistency assessment across regions. To address this challenge, we propose novel estimators and inference methods utilizing inverse probability of sampling and calibration weighting. Our approaches aim to eliminate exogenous regional imbalance while preserving intrinsic differences across regions, such as race and genetic variants. Moreover, time-to-event outcomes in MRCT studies receive limited attention, with existing methodologies primarily focusing on hazard ratios. In this paper, we adopt restricted mean survival time to characterize the treatment effect, offering more straightforward interpretations of treatment effects with fewer assumptions than hazard ratios. Theoretical results are established for the proposed estimators, supported by extensive simulation studies. We illustrate the effectiveness of our methods through a real MRCT case study on acute coronary syndromes.
\end{abstract}

\noindent%
\vfill

\newpage
\spacingset{1.9} 

\section{Introduction}  \label{section1}
\vspace*{-2ex}

\subsection{Multi-Regional Clinical Trials} \label{section1.1}
With a growing demand for region-specific evidence of drug effects and a need for pooling patient data from multiple regions, the pharmaceutical industry has globalized the research and development to provide high-quality medical products around the world~\citep{quan2013empirical}. Multi-regional clinical trials (MRCTs), in which patients are enrolled from multiple geographically separated regions, have become a common practice in recent years. They could expedite the availability of critical medical products to patients globally and enhance developmental efficiency in regional clinical research~\citep{chen2010assessing}. In 2016, the International Council for Harmonisation E17~\citep{guideline2016general} was issued to promote MRCT data acceptance by multiple regulatory agencies.

In MRCT, we aim to estimate region-specific treatment effects and assess their consistency across regions. 
Consistent treatment effects allow for data combination, which enhances the power to detect overall treatment effects and extends the generalizability of trial findings. However, in the presence of inconsistency in treatment effects and safety profiles, mainly if the differences are substantial or opposite, regional regulatory agencies may opt to disapprove the drug for their market.
\vspace*{-2ex}

\subsection{Essential and Inessential Traits in MRCT} \label{section1.2}
In MRCT, the treatment effect inconsistency between regions can stem from various factors that can be classified into two categories: \emph{exogenous (inessential)} imbalance and \emph{intrinsic (essential)} imbalance of regional traits. Essential traits are intrinsic to a region, including a region's ethnic and racial constituents, genetic variants, and medical practice due to administrative, cultural, and historical reasons. These factors are typically important to distinguish regions but may be difficult to quantify and often not captured by the MRCTs. In contrast, inessential traits, such as patient characteristics and clinical factors, are exogenous to a region's identity. Discrepancies in these inessential traits among regions are inevitable due to patient recruitment patterns, thus partly contributing to variations in treatment effects across regions.

With the obligation to protect the patients of their regions from ineffective or toxic treatments, regulatory agencies are usually interested in the inconsistency caused by essential traits. However, assessing treatment effect consistency across regions may be biased if imbalanced inessential traits modify the treatment effect. Therefore, it is crucial to eliminate any apparent difference introduced by these inessential traits, particularly effect modifiers, within a specific sample to ensure a comparable evaluation of treatment effects. Throughout this paper, we will refer to inessential traits as variables that also modify the region-specific average treatment effect (defined in Section~\ref{section2}), and we will interchangeably use the terms ``inessential traits'' and ``covariates".
\vspace*{-2ex}

\subsection{Motivation Example: PLATO Trial} \label{section1.3}
This paper is motivated by the PLATO (PLATlet inhibition and patient Outcomes) trial\\~\citep{wallentin2009ticagrelor}, which compared a novel antiplatelet inhibitor, ticagrelor, versus a standard antiplatelet therapy, clopidogrel for patients with acute coronary syndromes. Conducted across 43 countries with 18,624 patients, the trial's primary endpoint was the time to cardiovascular death, myocardial infarction, or stroke. Previous analysis~\citep{carroll2013statistical} revealed a treatment effect inconsistency between patients from the United States (US) and rest of the world (non-US), with ticagrelor showing more efficacy than clopidogrel in non-US but the opposite in the US. This inconsistency was mainly attributed to variations in aspirin dosage between two regions, and ticagrelor was found to be associated with a lower risk of the primary outcome compared to clopidogrel in patients taking low-dose aspirin~\citep{mahaffey2011ticagrelor}.

In our case study (Section~\ref{section6}), we propose to evaluate region-specific treatment effects in the PLATO trial by balancing the maintenance aspirin dosage and other inessential traits across regions. However, the existing methods in MRCT ignore the incomparability of patients' covariates from different regions under fixed or random effects hierarchical models~\citep{chen2010assessing,tsong2012assessment,quan2013empirical,quan2014multi}. They may lead to biased estimates of region-specific treatment effects and inaccurate consistency assessments. To address this limitation, we propose advanced approaches to eliminate disparities arising from inessential traits, providing a more robust and unbiased assessment of region-specific treatment effects in MRCTs.
\vspace*{-2ex}

\subsection{Objectives} \label{section1.4}
To eliminate the disparities arising from inessential traits, we propose to generalize the treatment effect from each region to a target distribution of these covariates. By doing so, we can ascribe any variations in the generalized region-specific treatment effects to the essential traits. Many generalization methods rely on inverse probability of sampling weighting (IPSW) approaches~\citep{stuart2011use,kern2016assessing,westreich2017transportability,dahabreh2019generalizing,dahabreh2020extending}. These approaches rely on correct specification of the sampling model and sufficient overlap between study participants and target population~\citep{degtiar2023review}. However, they become unstable due to model misspecification, extreme weights, or sparse covariates~\citep{robins2007comment,ben2021balancing}. Furthermore, IPSW-based approaches are ineffective in cases where individual patient data from the target population is not accessible~\citep{chattopadhyay2022one}.

To address these issues from IPSW, recent research embraced the calibration weighting (CW) method to generalize treatment effects from clinical trials to the target population~\citep{josey2021transporting,josey2022calibration,lee2021improving}. The calibration weights are estimated by solving an entropy maximization problem under constraints of an exact balance of covariate moments~\citep{hainmueller2012entropy,zhao2016entropy,wang2019bias,lee2021improving} so that the covariate distributions of the samples across groups empirically match the common target population. Estimating the calibration weights does not require fitting sampling or outcome models. In addition, this method is flexible as it is applicable when only the sample moments of the covariates in the target population are available~\citep{josey2021transporting}.

In this paper, we propose both IPSW and CW-based methods to balance the covariate distributions across regions in an MRCT against a common target population that carries clinical relevance and interpretable validity. Although IPSW-based approaches have limitations, they have not been extensively discussed in the context of MRCTs. We focus on the CW-based treatment effect estimators, enabling us to consistently estimate the average region-specific treatment effects concerning the target population. Furthermore, we assess the treatment effect consistency across regions and estimate a global average treatment effect if consistency holds. Our approach can eventually enhance the interpretability of MRCTs.

While previous studies have primarily concentrated on binary outcomes, limited attention has been given to time-to-event outcomes. To fill this methodological and practical gap, we propose inference methods for MRCTs with time-to-event endpoints. We incorporate the restricted mean survival time in our proposed methods, leveraging the advantages of RMST in analyzing time-to-event data in clinical studies~\citep{royston2013restricted,uno2014moving,tian2014predicting,perego2020utility}.

The remainder of this paper is structured as follows. Section~\ref{section2} establishes the basic settings and assumptions for the problem. 
Section~\ref{section3} introduces the IPSW and CW methods in MRCT. In Section~\ref{section4}, we propose four weighted estimators for region-specific average RMST difference and introduce a regional consistency test of treatment effects. In Section~\ref{section5}, we show an extensive simulation study and assess the finite sample properties of the proposed estimators. The case study in Section~\ref{section6} illustrates our methods on the PLATO trial. We conclude the paper with discussions and future works in Section~\ref{section7}.
\vspace*{-2ex}

\section{Basic Setup} \label{section2}
\vspace*{-2ex}
\subsection{Notations and MRCT Data Structure} \label{section2.1}
\begin{figure}[!htbp]
\centering\includegraphics[scale = 0.6]{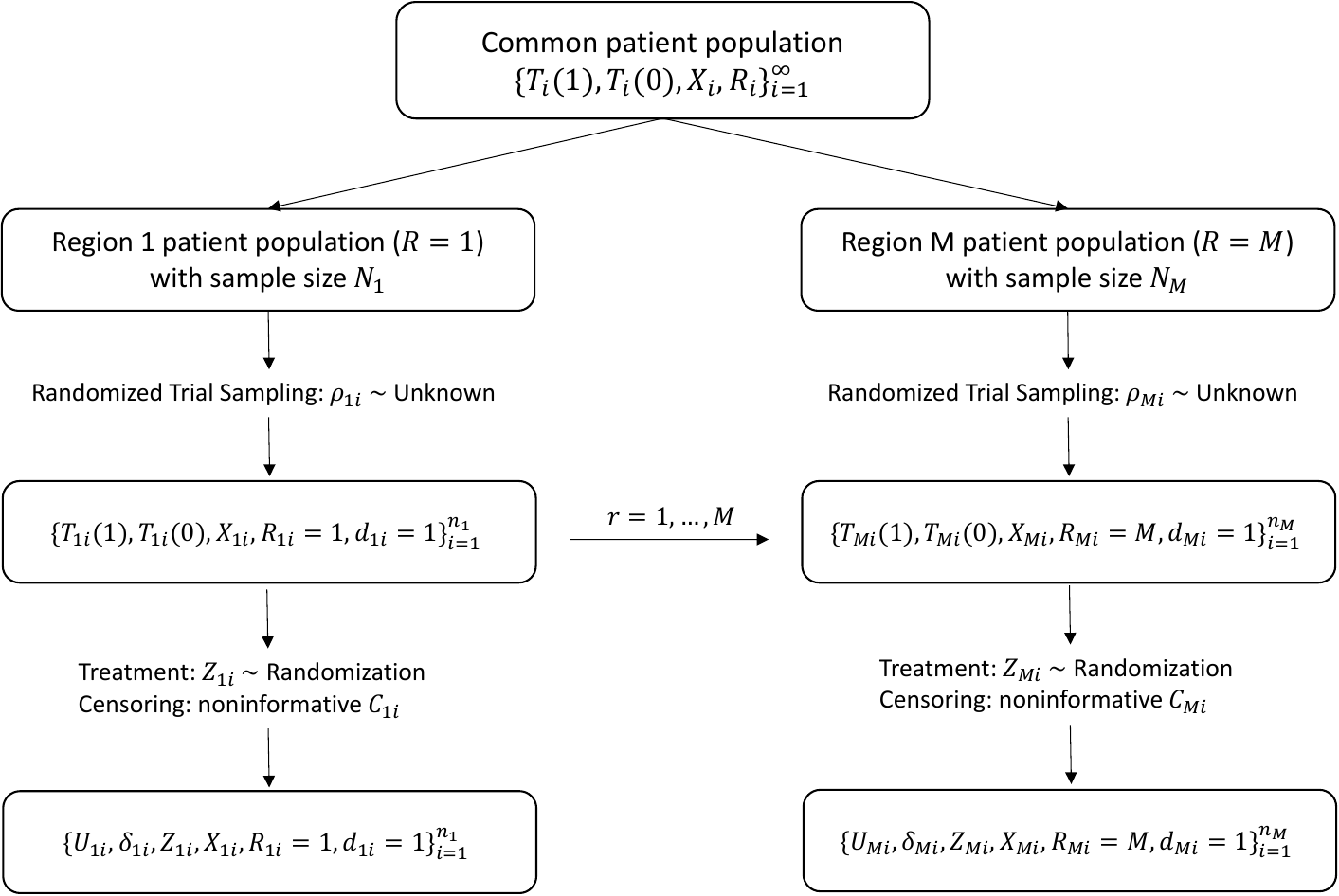}
\caption{Data structure of multi-regional clinical trials.}
\label{fig1}
\end{figure}

Figure~\ref{fig1} displays the MRCT data structure. Suppose there are $M$ geographically separate regions ($M \geq 2$) with $R$ as a region indicator ($R=1,\dots,M$). Let $N_r$ denote the size of the patient population with a certain disease in Region $r$ ($R=r$) under review of the enrollment for a randomized trial, which is not necessarily known. Hypothetically, we assume the patient population from different regions share a common distribution of inessential traits (e.g., gender and age), while they may be differentiate by the essential traits (e.g., race). Suppose $\bX$ is a $p$-dimensional covariate vector of inessential traits, and $F$ denotes the common distribution of $\bX$ (i.e., $X \sim F(\bX)$).

We first consider a randomized trial in a single Region $r$, comparing the efficacy of two treatments. Out of $N_r$ individuals, suppose $n_r$ patients are enrolled in the trial. We indicate the selection using $d$, where $d=1$ if a patient is enrolled in the trial and $d=0$ otherwise. In general, we can only observe the data of the enrolled patients (i.e., $d=1$). Suppose $\rho_r(\bX) = P(d=1|\bX,R=r)$ is an unknown sampling score for trial participation in Region $r$ given $\bX$. We denote the distribution of $\bX$ for the trial participants as $F_r(\bX) = F(\bX|d=1,R=r)$. Let $Z$ be the indicator of the treatment assignment, where $Z=1$ for the treatment group and $Z=0$ for the control group. Let $\pi_r(\bX) = P(Z=1|\bX,R=r,d=1)$ be the treatment propensity score. In randomized trials, $\pi_r(\bX) = \pi_r$ can be commonly assumed to be a constant and known by design.

Following the potential outcomes framework in~\cite{rubin1974estimating}, let $T(1)$ and $T(0)$ denote the potential time-to-event under treatment and control, respectively. By assuming the stability of potential outcomes, the time-to-event is $T = T(1)Z + T(0)(1-Z)$. Let $C$ denote the censoring time. In the presence of right censoring, we observe $U = \text{min}(T,C)$ and the censoring indicator $\delta = I[T \leq C]$, where $I[\cdot]$ is an indicator function. In summary, let $i$ index the enrolled patient, the observed data in Region $r$ is $\{U_{ri}, \delta_{ri}, \bX_{ri}, Z_{ri}, R_{ri}=r, d_{ri}=1 | i = 1,\dots,n_r\}$, where $r = 1,\dots,M$.
\vspace*{-2ex}

\subsection{Restricted Mean Survival Time} \label{section2.2}
Restricted mean survival time (RMST)~\citep{irwin1949standard,andersen2004regression,royston2013restricted,uno2014moving} summarizes the survival time up to a clinically relevant (usually pre-specified) truncation time $t^*$. It is defined as the mean of the truncated event time $Y = \text{min}(T, t^*)$, and can be calculated by the area under the survival curve $S(t) = P[T>t]$ from $t = 0$ to $t = t^*$~\citep{royston2013restricted}:
\[
\mu(t^*) = E(Y) = E[T \wedge t^*] = \int_0^{t^*}S(t)dt.
\] 
The difference or ratio of RMST between two treatments measures the relative treatment effect concerning a gain or loss of event-free survival time up to $t^*$~\citep{kim2017restricted}. Compared to the hazard ratio, typically estimated from the Cox proportional hazard (PH) model~\citep{cox1972regression}, which can be a misleading and inappropriate summary of treatment effect due to PH violations~\citep{lin1989robust}, estimating RMST does not require model assumptions. RMST measures offer more straightforward interpretations than hazard ratios across various distributions of time-to-event outcomes~\citep{perego2020utility}.

\subsection{Estimands and Assumptions} \label{section2.3}
In Region $r$, we define the treatment-specific conditional survival function as \\
$S_{rz}(t|\bX) = E\{I[T(z) \geq t]|\bX,R=r,d=1\}$ for $z \in \{0,1\}$. For a truncation time $t^*$, the treatment-specific conditional RMST up to $t^*$ is $\mu_{rz}(t^*|\bX) = \int_0^{t^*}S_{rz}(t|\bX)dt$. The conditional treatment effect is defined by the RMST difference such as $\Delta_{r}(t^*|\bX) = \mu_{r1}(t^*|\bX) - \mu_{r0}(t^*|\bX)$.

We seek to generalize the region-specific conditional RMST difference $\Delta_{r}(t^*|\bX)$ to a target population with a covariate distribution of $\bX \sim F^*(\bX)$. Our estimand is a region-specific average RMST difference if the distribution of covariates $\bX$ from the trial participants in Region $r$ is the same as in the target population, which is written as: 
\begin{eqnarray} \label{eqn2.1}
    \Delta_{r}(t^*) = \mu_{r1}(t^*) - \mu_{r0}(t^*),
\end{eqnarray}
where $\mu_{rz}(t^*) = E_{F^*}[\mu_{rz}(t^*|\bX)]$ is the region-specific average RMST for $z \in \{0,1\}$, with expectation taken on $F^*(\bX)$. To identify the estimand from observed data in MRCT, we make the following assumptions:

\begin{assumption}[Ignorability and positivity of trial treatment assignment] \label{assumption1}
(i) $\{T(1),T(0)\} \ind Z| (d=1,\bX,R)$, and (ii) $0 < \pi_r(\bX)< 1$ with probability 1.
\end{assumption}

\begin{assumption}[Conditional noninformative censoring]
\label{assumption2}
$\{T(1),T(0)\} \ind C | (d=1,Z,\bX,R)$, which implies $T \ind C | (d=1,Z,\bX,R)$.
\end{assumption}

\begin{assumption}[Covariate overlap with target population]
\label{assumption3}
The target distribution of $\bX$, $F^*$, is absolutely continuous concerning the distribution of $\bX$ for trial participants in each region, $F_r$. That is, for any set of $\bX$, $A_{\bX}$, if $A_{\bX}$ has zero probability in $F_r$, then it also has zero probability in $F^*$. 
\end{assumption}

\begin{assumption}[Conditional exchangeability of survival function and positivity of trial participation]
\label{assumption4}
(i) $S(t|\bX,Z,R,d=1) = S(t|\bX,Z,R)$, and (ii) $\rho_r(\bX) > 0$ with probability 1.
\end{assumption}

Assumption~\ref{assumption4} also implies the conditional exchangeability of RMST, i.e., $\mu_{rz}(t^*|\bX) = \int_0^{t^*}S(t|\bX,Z=z,R=r,d=1)dt = \int_0^{t^*}S(t|\bX,Z=z,R=r)dt$. Under Assumptions~\ref{assumption1} to~\ref{assumption4}, $\mu_{rz}(t^*)$ can be identified by $E[\mu_{rz}(t^*|\bX)\frac{dF^*}{dF_r}(\bX)]$, where $\frac{dF^*}{dF_r}(\bX)$ denotes the Radon-Nikodym derivative of the distribution $F^*$ relative to the distribution of $F_r$. For example, if all covariates in $\bX$ are continuous, $\frac{dF^*}{dF_r}(\bX)$ is the ratio of the probability density functions.
\vspace*{-2ex}

\section{Weighting Methods in MRCT} \label{section3}
\vspace*{-2ex}

\subsection{The Inverse Probability of Sampling Weights} \label{section3.1}
Following the concept of the balancing weight discussed in~\cite{li2018balancing}, to generalize the treatment effect from each region to the target population, we can weigh the distribution of $F_r$ to the target population $F^*$ by using the following weight functions:
\begin{eqnarray} \label{eqn3.1}
\gamma_{ri} = \frac{dF^*(\bX_{ri})}{dF_r(\bX_{ri})} = \frac{P_{F^*}(\bX = \bX_{ri})}{P_{F_r}(\bX = \bX_{ri})},
\end{eqnarray}
where $P_{F^*}$ and $P_{F_r}$ denote the joint probability of $\bX = \bX_{ri}$ under the distribution of $F^*$ and $F_r$, respectively. However, as the density functions may be difficult to estimate when $\bX$ is high-dimensional~\citep{westreich2017transportability}, we can use Bayes' rules to rearrange the $\gamma_{ri}$ from Equation~\eqref{eqn3.1} into an inverse probability of sampling weight (IPSW) as follows:
\begin{eqnarray} \label{eqn3.2}
\gamma_{ri} = \frac{P(d'_{ri}=1|\bX_{ri})}{P(d_{ri}=1|\bX_{ri},R_{ri}=r)}\frac{P(d_{ri}=1|R_{ri}=r)}{P(d'_{ri}=1)} \propto \frac{P(d'_{ri}=1|\bX_{ri})}{P(d_{ri}=1|\bX_{ri},R_{ri}=r)},
\end{eqnarray}
where $d'_{ri} \in \{0,1\}$ denotes the inclusion in the target population. As a result, the weight $\gamma_{ri}$ from Equation~\eqref{eqn3.2} is proportional to the inverse ratio of a patient's sampling score for trial participation in Region $r$ as opposed to the probability of being in the target population, conditional on the covariates $\bX_{ri}$. However, this approach requires prior knowledge of the sampling score of the target population, $P(d'_{ri}=1|\bX_{ri})$.

According to the MRCT data structure shown in Figure~\ref{fig1}, we can only access the data from the trial participation population in each region (i.e., $d=1$), and the data from other unenrolled patients are unavailable. Therefore, we may not directly estimate the sampling score $\rho_r(\bX) = P(d=1|\bX,R=r)$ in each region only with the data of $d=1$. Alternatively, one may estimate the sampling score by using a region-specific propensity score, $P(R=r|\bX,d=1)$, given the accessible data. One option is to use the gradient boosted model~\citep{mccaffrey2013tutorial,burgette2021propensity} to estimate $P(R=r|\bX,d=1)$. This method is implemented using the R package \texttt{twang} (version 1.4-9.5)~\citep{ridgeway2017package}. However, using a misspecified sampling score model may lead to biased results when implementing the IPSW. In the simulation study (Section~\ref{section5}), we will show that the IPSW method is sensitive to the model specification of $\rho_r(\bX)$.

\subsection{The Calibration Weights} \label{section3.2}
In contrast, the calibration weights (CW) can be estimated without fitting the sampling score model. We utilize the entropy balancing method proposed by~\cite{hainmueller2012entropy} to estimate the calibration weights. Let $p_{ri}$ denote the weights, where $i$ and $r$ index the patient and region such as $i = 1,\dots,n_r$ and $r = 1,\dots,M$. We estimate $p_{ri}$ by solving the following optimization problem:
\begin{equation} \label{eqn3.3}
    \text{min} \left \{ \sum_{r=1}^{M}\sum_{i=1}^{n_r}p_{ri}\text{log}(p_{ri}) \right \},
\end{equation}
with constraints:
\begin{eqnarray}
\sum_{i=1}^{n_r}p_{ri}\bg(\bX_{ri}) &=& \Tilde{\bg}, r = 1,\dots,M. \label{eqn3.4}\\
\sum_{i=1}^{n_r}p_{ri} &=& 1, r = 1,\dots,M. \label{eqn3.5}
\end{eqnarray}
The objective function in Equation~\eqref{eqn3.3} is the negative entropy of the CWs. Minimizing this function keeps the CW's empirical distribution close to the uniform, which in turn minimizes the variability due to heterogeneous weights~\citep{lee2021improving}.

In Equation~\eqref{eqn3.4}, we suppose $\bg(\cdot)$ contains $L$ covariate functions to be calibrated. Let $\bg(\bX) = [g_1(\bX),\dots,g_L(\bX)]$, each $g_l(\cdot)$ is a function of the covariates that can be any transformation of $\bX$, but not necessarily the polynomial function~\citep{zhao2016entropy}. We impose these functions to equalize the moments of the covariates between each region and the target population. A typical balance constraint contains the first $k^{th}$ moments of the calibrated covariates~\citep{hainmueller2012entropy,josey2021framework}. For example, when $k=2$, $\bg(\bX) = [\bX, \bX^2]$, which forces the first and second moment of $\bX$ to equal each region and the target population. In general, lower order $k$ (e.g., $k \leq 2$) is often sufficient to empirically match the covariate distributions from trial participants in every region to the target population~\citep{signorovitch2010comparative,wang2019bias}. We let $\Tilde{\bg} = [\Tilde{g}_1,\dots,\Tilde{g}_L]$ be the associated moment estimates of $\bg(\bX)$ from any researcher-defined target population.

Equation~\eqref{eqn3.5} implies that the calibration weights $p_{ri}$ in each region sum to a normalization constant of one, which guarantees that $p_{ri}$ are valid density functions in each region~\citep{hainmueller2012entropy,zhao2016entropy}.

We calculate $\hat{p}_{ri}$ in Equation~\eqref{eqn3.3} by using the Lagrange multiplier~\citep{de2000mathematical}.
\[
\hat{p}_{ri} = \frac{\text{exp}\{\blambda_r^T\bg(\bX_{ri})\}}{\sum_{i=1}^{n_r}\text{exp}\{\blambda_r^T\bg(\bX_{ri})\}},
\]
where $\blambda_r$ solves $\sum_{i=1}^{n_r}\text{exp}\{\blambda_r^T\bg(\bX_{ri})\}\{\bg(\bX_{ri}) - \Tilde{\bg}\} = 0$, for $r = 1,\dots,M$. We provide the derivation in the Supplementary Materials (Web Appendix A). 
\vspace*{-2ex}


\section{Weighted Estimators for Region-Specific RMST Difference} \label{section4}
We propose four weighted estimators for the region-specific RMST difference in MRCT. It is important to note that each proposed estimator can employ both weighting methods as introduced in Section~\ref{section3}. Throughout this section, we denote $\hat{\xi}_{ri}$ as the estimated weight function that can be either IPSW $\hat{\gamma}_{ri}$ or CW $\hat{p}_{ri}$. In the remainder of this paper, we denote ``CW-adjusted estimators" and ``IPSW-adjusted estimators" as the proposed weighted estimators using $\hat{p}_{ri}$ and $\hat{\gamma}_{ri}$, respectively.

We derive the large sample properties of each proposed estimator, and some additional assumptions are required. For the IPSW-adjusted estimators, we assume the sampling score model $\rho_r(\bX)$ is correctly specified. For the CW-adjusted estimators, we assume that for any uncalibrated covariates, denoted as $\bW$, the sampling score is conditionally exchangeable, i.e., $P(d=1|\bX,\bW,R=r) = P(d=1|\bX,R=r) = \rho_r(\bX)$. The proofs of all theoretical results are given in the Supplementary Materials (Web Appendix B to D).

\subsection{Weighted Kaplan-Meier (KM) RMST Difference} \label{section4.1}
The first estimator is derived based on a weighted RMST estimator proposed by~\cite{conner2019adjusted} by integrating a weighted KM survival curve proposed by~\cite{xie2005adjusted}. Based on their work, we propose a weighted KM estimator for the region-specific average RMST difference in MRCT. For $i=1,\dots,n_r$, denote $N_{rzi}(t) = I[U_{ri} \leq t; \delta_{ri}=1; Z_{ri}=z]$ as the individual treatment-specific counting process and $Y_{rzi}(t) = I[U_{ri} \geq t; Z_{ri}=z]$ as the individual treatment-specific at risk process. Then for treatment $z \in \{0,1 \}$, the weighted KM RMST estimator for $\mu_{rz}(t^*)$ is written as:
\begin{equation} \label{eqn4.1}
\Tilde{\mu}_{rz}(t^*) = \int_0^{t^*}\Tilde{S}_{rz}(t)dt = \int_0^{t^*} \prod_{u \leq t} \left \{ 1 - \frac{d\Tilde{N}_{rz}(u)}{\Tilde{Y}_{rz}(u)}\right \} dt,
\end{equation}
where $\Tilde{S}_{rz}(t)$ is a weighted KM estimator for the region-specific average survival function. We define $\Tilde{N}_{rz}(u) = \sum_{i=1}^{n_{r}}\hat{\xi}_{ri}q_{rzi}(\bX_{ri})N_{rzi}(u)$ as the weighted counting process and $\Tilde{Y}_{rz}(u) = \sum_{i=1}^{n_{r}}\hat{\xi}_{ri}q_{rzi}(\bX_{ri})Y_{rzi}(u)$ as the weighted at-risk process. Here, $q_{rzi}(\bX_{ri}) = \\ \frac{I(Z_{ri} = z)}{\pi_{ri}(\bX_{ri})^z(1 - \pi_{ri}(\bX_{ri}))^{1-z}}$ is an inverse of the individual treatment-specific propensity score, where $\pi_{ri}(\bX_{ri})$ is the treatment propensity score that can be either estimated or assumed as constant.

The variance of $\Tilde{\mu}_{rz}(t^*)$ is derived as follows:
\begin{equation} \label{eqn4.2}
\mbox{Var}[\Tilde{\mu}_{rz}(t^*)] 
= \int_0^{t^*} \left \{ \int_u^{t^{*}}\Tilde{S}_{rz}(t)dt \right \}^2 \frac{d\Tilde{N}_{rz}(u)}{\Tilde{W}_{rz}(u)(\Tilde{Y}_{rz}(u) - \Delta \Tilde{N}_{rz}(u))},
\end{equation}
where $\Tilde{W}_{rz}(u) = \frac{\Tilde{Y}_{rz}^2(u)}{\sum_{i=1}^{n_r}[\hat{\xi}_{ri}q_{rzi}(\bX_{ri})]^2Y_{rzi}(u)}$ and $\Delta \Tilde{N}_{rz}(u) = \Tilde{N}_{rz}(u) - \Tilde{N}_{rz}(u-)$. Note that if we assume that the treatment propensity score $\pi_{ri}(\bX_{ri})$ is a constant, the term $q_{rzi}(\bX_{ri})$ can be omitted when calculating $\Tilde{\mu}_{rz}(t^*)$ and its variance.

The weighted KM estimator for $\Delta_r(t^*)$ is then defined as:
\begin{eqnarray} \label{eqn4.3}
    \hat{\Delta}_r^{KM}(t^*) = \Tilde{\mu}_{r1}(t^*) - \Tilde{\mu}_{r0}(t^*),
\end{eqnarray}
and we have the following theoretical properties for $\hat{\Delta}_r^{KM}(t^*)$:
\begin{theorem}
For any fixed $0 < t^* < \infty$, as $n_r \to \infty$,
\begin{eqnarray*}
\left |\hat{\Delta}_r^{KM}(t^*) - \Delta_r(t^*) \right | &\overset{p}{\to}& 0, \\
\sqrt{n_r}\left\{ \hat{\Delta}_r^{KM}(t^*) - \Delta_r(t^*) \right\} &\overset{d}{\to}& N \left ( 0,\sigma_{r,KM}^2(t^*) \right ).
\end{eqnarray*}
Here, by $\overset{p}{\to}$ and $\overset{d}{\to}$ we mean ``converges in probability" and ``converges in distribution", respectively. Let $\Lambda_{rz}(t) = E_{\bX}[-log \{ S(t|\bX, Z = z, R = r, d = 1) \} ]$ denote the cumulative incidence function, the asymptotic variance \\
$\sigma_{r,KM}^2(t^*) = n_r\sum_{z=0,1}\int_0^{t^*}\{\int_u^{t^*}S_{rz}(t)dt\}^2\frac{d\Lambda_{rz}(u)}{\Tilde{W}_{rz}(u)}$, can be estimated by:
\begin{eqnarray*}
\hat{\sigma}_{r,KM}^2(t^*) = n_r\sum_{z=0,1}\int_0^{t^*} \left \{ \int_u^{t^{*}}\Tilde{S}_{rz}(t)dt \right \}^2 \frac{d\Tilde{N}_{rz}(u)}{\Tilde{W}_{rz}(u)(\Tilde{Y}_{rz}(u) - \Delta \Tilde{N}_{rz}(u))}.
\end{eqnarray*}
\end{theorem}

\subsection{Weighted G-Formula (GF) RMST Difference} \label{section4.2}
The second estimator employs the G-computation technique~\citep{robins1986new,robins2008estimation,naimi2017introduction}. This estimator is a direct regression estimator, and its outcome model can be identified by the inverse probability of censoring weighted (IPCW) RMST regression~\citep{tian2014predicting}.

For a specific truncation time $t^*$, we define the conditional RMST up to $t^*$ given the covariates and treatment as $\mu(t^{*}|\bX,Z) = E(Y|\bX,Z)$. In each Region $r$, we fit an IPCW RMST regression model as follows:
\begin{eqnarray} \label{eqn4.4}
    \phi(\mu_r(t^*|\bX_r,Z_r)) = \beta_{r0} + \beta_{r1}Z_r + \bbeta_{r2}\bg(\bX_r^T) + \bbeta_{r3} Z_r\bg(\bX_r^T),
\end{eqnarray}
where $\phi(\cdot)$ is a link function, and log or identify links are often used. Here, $\bg(\cdot)$ is the same function mentioned in Equation~\eqref{eqn3.4}. Based on Equation~\eqref{eqn4.4}, we define the outcome models for $Z_r \in \{0,1\}$ in Region $r$ as $m_{r0}(\bX_r) = \phi^{-1}(\hat{\beta}_{r0} + \hat{\bbeta}_{r2}\bg(\bX_r^T))$ and $m_{r1}(\bX_r) = \phi^{-1}(\hat{\beta}_{r0} + \hat{\beta}_{r1} + \hat{\bbeta}_{r2}\bg(\bX_r^T) + \hat{\bbeta}_{r3}\bg(\bX_r^T))$, respectively. Then, the weighted GF estimator for $\Delta_r(t^*)$ is written as:
\begin{eqnarray}  \label{eqn4.5}
    \hat{\Delta}_r^{GF}(t^*) = \frac{\sum_{i=1}^{n_r} \hat{\xi}_{ri}\{m_{r1}(\bX_{ri}) - m_{r0}(\bX_{ri})\}}{\sum_{i=1}^{n_r} \hat{\xi}_{ri}}.
\end{eqnarray}
By assuming that the IPCW RMST regression model in Equation~\eqref{eqn4.4} is not misspecified, we have the following theoretical properties for $\hat{\Delta}_r^{GF}(t^*)$:
\begin{theorem}
For any fixed $0 < t^* < \infty$, as $n_r \to \infty$,
\begin{eqnarray*}
\left |\hat{\Delta}_r^{GF}(t^*) - \Delta_r(t^*) \right | &\overset{p}{\to}& 0, \\
\sqrt{n_r}\left\{ \hat{\Delta}_r^{GF}(t^*) - \Delta_r(t^*) \right\} &\overset{d}{\to}& N \left ( 0,\sigma_{r,GF}^2(t^*) \right ).
\end{eqnarray*}
Here, the asymptotic variance $\sigma_{r,GF}^2(t^*)$ can be estimated using the Delta method~\citep{dowd2014computation}:
\begin{eqnarray*}
    \hat{\sigma}_{r,GF}^2(t^*) = n_r J_r^T \Sigma_r J_r,
\end{eqnarray*}
where $\Sigma_r$ is the variance-covariance matrix of the parameter vector $\hat{\bbeta}_r=[\hat{\beta}_{r0},\hat{\beta}_{r1},\hat{\bbeta}_{r2},\hat{\bbeta}_{r3}]^T$ and $J_r = \partial \hat{\Delta}_r^{GF}(t^*)/\partial \hat{\bbeta}_r$.
\end{theorem}

\subsection{Weighted Hajek (HJ) RMST Difference} \label{section4.3}
The third estimator is based on a RMST estimator using the IPCW~\citep{bang2000estimating}. In a general context, suppose there are $n$ patients and let $Y_i = min(T_i,t^*)$ for $i=1,\dots,n$. In the presence of right censoring, the IPCW estimator for $\mu(t^*)$ is
\begin{eqnarray} \label{eqn4.6}
    \hat{\mu}_{IPCW}(t^*) = \frac{1}{n}\sum_{i=1}^n\frac{\delta_i^*}{\hat{G}(Y_i)}Y_i,
\end{eqnarray}
where $\delta_i^* = I[C_i \geq Y_i]$ and $\hat{G}(Y_i) = P(C_i > Y_i)$ is the KM estimator of the survival function for the censoring time based on $\{(U_i, 1-\delta_i), i=1,\dots,n\}$. To understand the validity of $\hat{\mu}_{IPCW}(t^*)$, note that
\begin{eqnarray*}
    E \left[ \frac{\delta_i^*}{\hat{G}(Y_i)}Y_i | T_i \right] = Y_i\frac{P(C_i \geq Y_i|T_i)}{P(C_i > Y_i)} = Y_i.
\end{eqnarray*}

Based on Equation~\eqref{eqn4.6}, the weighted HJ estimator for $\Delta_r(t^*)$ is defined as:
\begin{eqnarray} \label{eqn4.7}
    \hat{\Delta}_r^{HJ}(t^*) = \frac{\sum_{i=1}^{n_r}\hat{\xi}_{ri}q_{r1i}(\bX_{ri})w_{ri}Y_{ri}}{\sum_{i=1}^{n_r}\hat{\xi}_{ri}q_{r1i}(\bX_{ri})w_{ri}} - \frac{\sum_{i=1}^{n_r}\hat{\xi}_{ri}q_{r0i}(\bX_{ri})w_{ri}Y_{ri}}{\sum_{i=1}^{n_r}\hat{\xi}_{ri}q_{r0i}(\bX_{ri})w_{ri}},
\end{eqnarray}
where $w_{ri} = \delta_{ri}^*/\hat{G}(Y_{ri})$ with $\delta_{ri}^* = I[C_{ri} \geq Y_{ri}]$ and $\hat{G}_r(Y_{ri}) = P(C_{ri} > Y_{ri})$. We have the following theoretical properties for $\hat{\Delta}_r^{HJ}(t^*)$:
\begin{theorem}
For any fixed $0 < t^* < \infty$, as $n_r \to \infty$,
\begin{eqnarray*}
\left |\hat{\Delta}_r^{HJ}(t^*) - \Delta_r(t^*) \right | &\overset{p}{\to}& 0, \\
\sqrt{n_r}\left\{ \hat{\Delta}_r^{HJ}(t^*) - \Delta_r(t^*) \right\} &\overset{d}{\to}& N \left ( 0,\sigma_{r,HJ}^2(t^*) \right ).
\end{eqnarray*}
\end{theorem}
\noindent This theorem can be proven by the M-estimator theory~\citep{stefanski2002calculus}, and we provide the derivation of the asymptotic variance $\sigma_{r,HJ}^2(t^*)$ in the Supplementary Materials (Web Appendix D).

\subsection{Weighted Augmented (AG) RMST Difference} \label{section4.4}
The weighted Augmented RMST estimator combines the weighted G-formula and Hajek estimators:

\begin{eqnarray}
    \hat{\Delta}_r^{AG}(t^*) &=& 
    \frac{\sum_{i=1}^{n_r}\hat{\xi}_{ri}q_{r1i}(\bX_{ri})w_{ri}\{Y_{ri}-m_{r1}(\bX_{ri})\}}{\sum_{i=1}^{n_r}\hat{\xi}_{ri}q_{r1i}(\bX_{ri})w_{ri}} \nonumber \\
    &-& \frac{\sum_{i=1}^{n_r}\hat{\xi}_{ri}q_{r0i}(\bX_{ri})w_{ri}\{Y_{ri}-m_{r0}(\bX_{ri})\}}{\sum_{i=1}^{n_r}\hat{\xi}_{ri}q_{r0i}(\bX_{ri})w_{ri}}  \nonumber \\ 
    &+& \frac{\sum_{i=1}^{n_r} \hat{\xi}_{ri}\{m_{r1}(\bX_{ri}) - m_{r0}(\bX_{ri})\}}{\sum_{i=1}^{n_r} \hat{\xi}_{ri}}.
\end{eqnarray}
Based on the semiparametric theory~\citep{tsiatis2006semiparametric}, $\hat{\Delta}_r^{AG}(t^*)$ is doubly robust and does not require the correct specification of the outcome models (shown in simulation study and Supplementary Materials). The theoretical properties for $\hat{\Delta}_r^{AG}(t^*)$ is as follows:
\begin{theorem}
For any fixed $0 < t^* < \infty$, as $n_r \to \infty$,
\begin{eqnarray*}
\left |\hat{\Delta}_r^{AG}(t^*) - \Delta_r(t^*) \right | &\overset{p}{\to}& 0, \\
\sqrt{n_r}\left\{ \hat{\Delta}_r^{AG}(t^*) - \Delta_r(t^*) \right\} &\overset{d}{\to}& N \left ( 0,\sigma_{r,AG}^2(t^*) \right ).
\end{eqnarray*}
\end{theorem}
\noindent This theorem is also proven by the M-estimator theory, and we provide the derivation of the asymptotic variance $\sigma_{r,AG}^2(t^*)$ in the Supplementary Materials (Web Appendix D).

\subsection{Regional Consistency Test} \label{section4.5}
We propose a Wald-type test to evaluate the consistency of treatment effects across regions. Given $t^*$, the null hypothesis of the treatment effect consistency is $H_{0}: \Delta_1(t^*)= \dots = \Delta_M(t^*)$. Regional consistency of treatment effects can be obtained if $H_0$ is not rejected. We define a $\chi^2$ test statistics $U(t^*)$ as follows:
\begin{eqnarray*}
\Tilde{U}(t^*) = \{ \bE\Tilde{\bDelta}(t^*)\}^T \{\bE\Tilde{\bV}(t^*)\bE^T\}^{-1} \bE\Tilde{\bDelta}(t^*),
\end{eqnarray*}
where $\bE$ is an $(M-1)$-by-$(M)$ contrast matrix
\text{\footnotesize
$\begin{pmatrix}
-1 & 1 & 0 & \dots & 0\\
-1 & 0 & 1 & \dots & 0\\
\dots & \dots & \dots & \dots & \dots\\
-1 & 0 & 0 & \dots & 1\\
\end{pmatrix}$ }, \\
$\Tilde{\bDelta}(t^*) = [\Tilde{\Delta}_1(t^*), \dots, \Tilde{\Delta}_M(t^*)]^T$, and $\Tilde{\bV}(t^*)$ is a diagonal matrix of $\mbox{Var}[\Tilde{\Delta}_r(t^*)]$, for $r = 1, \dots, M$. Here, $\Tilde{\Delta}_r(t^*)$ can be each of the estimators among $\hat{\Delta}_r^{KM}(t^*)$, $\hat{\Delta}_r^{GF}(t^*)$, $\hat{\Delta}_r^{HJ}(t^*)$, and $\hat{\Delta}_r^{AG}(t^*)$.
We assume the treatment effects are independent across regions. Under $H_0$, $\Tilde{U}(t^*)$ follows a $\chi^2$ distribution with $M-1$ degrees of freedom.

When the treatment effects are consistent across regions, we can combine the treatment effect estimates from each region and estimate a global treatment effect. The RMST difference for the global treatment effect is estimated by using an inverse variance weighted estimator~\citep{sinha2011statistical}:
\[
\Tilde{\Delta}_{G}(t^*) = \frac{\sum_{r=1}^M \Tilde{\Delta}_r(t^*)/\mbox{Var}[\Tilde{\Delta}_r(t^*)]}{\sum_{r=1}^M1/\mbox{Var}[\Tilde{\Delta}_r(t^*)]}.
\]
The variance of $\Tilde{\Delta}_{G}(t^*)$ is $\sum_{r=1}^M \frac{1}{\mbox{Var}[\Tilde{\Delta}_r(t^*)]} / \{ \sum_{r=1}^M\frac{1}{\mbox{Var}[\Tilde{\Delta}_r(t^*)]} \}^2$.

\section{Simulation Study} \label{section5}
\subsection{Aims and Performance Measures} \label{section5.1}
In this simulation study, we evaluate the finite sample performance of the proposed estimators for region-specific average RMST differences under various data-generating scenarios. The bias and variance of each estimator are evaluated under 1,000 Monte Carlo replications.

\subsection{Data-Generating Mechanism} \label{section5.2}
We simulate MRCTs conducted in three regions ($M=3$) and suppose the numbers of enrolled participants in each region are $n_1 = 400$, $n_2 = 500$, and $n_3 = 600$. We consider two covariates $\bX = [X_1, X_2]$ from a common population, where $X_1 \sim Unif(0,1)$ and $X_2 \sim N(1,1)$, and assume independence between $X_1$ and $X_2$. Suppose the distribution of $\bX$ in the target population is the same as the common population. We consider two types of sampling score models, the log-linear model, and the logistic model:
\begin{eqnarray}
    \text{log}\{\rho_r(\bX_{ri})\} &=& \eta_{r0} + \eta_{r1}X_{1ri} + \eta_{r2}X_{2ri}, \mbox{ for } r = 1,2,3. \label{eqn5.1} \\
    \text{logit}\{\rho_r(\bX_{ri})\} &=& \eta_{r0}^* + \eta_{r1}^*X^*_{1ri} + \eta_{r2}^*X^*_{2ri}, \mbox{ for } r = 1,2,3. \label{eqn5.2}
\end{eqnarray}
In Equation~\eqref{eqn5.1}, we assume the covariates have linear associations with the log of sampling score. In Equation~\eqref{eqn5.2}, we assume the covariates have non-linear associations with the logistic of sampling score by letting $X^*_{1ri} = X_{1ri} \times X_{2ri}$ and $X^*_{2ri} = \text{exp}\{X_{2ri}/10\}$. In each sampling score model, we consider two settings according to the level of similarity between enrolled patients in each region and the target population, which is quantified by the absolute standardized mean difference (SMD)~\citep{austin2011introduction} of covariates $\bX$. As a result, we have four scenarios as follows:
\begin{description}
    \item [Scenario 1:] Log-linear sampling with moderate SMDs.
    \item [Scenario 2:] Log-linear sampling with large SMDs.
    \item [Scenario 3:] Logistic-nonlinear sampling with moderate SMDs.
    \item [Scenario 4:] Logistic-nonlinear sampling with large SMDs.
\end{description}
These simulation scenarios evaluate the robustness of the proposed estimators under different sampling score models and similarities between each region and the target population. The actual values of the parameters $\eta_{r}$'s in Equation~\eqref{eqn5.1} and $\eta_r^*$'s in Equation~\eqref{eqn5.2}, and the absolute SMDs from each scenario are given in Web Tables 1 and 2 in the Supplementary Materials (Web Appendix E).

The following settings are equivalent in all four scenarios. We assume the treatment propensity score $\pi_r(\bX) = 0.5$ in all three regions, indicating the participants are one-to-one randomized to two treatment groups. The event time $T_{rzi}$ is assumed to be generated from the following hazard function:
\begin{eqnarray} \label{eqn5.3}
\begin{aligned}
&h(t|Z_{ri},X_{1ri},X_{2ri},R_{ri}) = \lambda(t|Z_{ri})\text{exp} \{ 0.3I[R_{ri}=2] + 0.5I[R_{ri}=3]\\ 
& - X_{1ri} + 0.5X_{2ri} + 0.3Z_{ri}I[R_{ri}=2] + 0.5Z_{ri}I[R_{ri}=3] \\
& - Z_{ri}X_{1ri} - 0.5Z_{ri}X_{2ri} - 0.6I[R_{ri}=2]X_{1ri} + 0.3I[R_{ri}=2]X_{2ri} \\
& - I[R_{ri}=3]X_{1ri} + 0.5I[R_{ri}=3]X_{2ri} \}.
\end{aligned}
\end{eqnarray}
We use different baseline hazard functions for each treatment group to simulate data under the non-PH assumption. We set the baseline hazard function for the experimental treatment group to be $\lambda(t|Z_{ri}=1) = 0.15t^{-0.7}$ and for the control group to be $\lambda(t|Z_{ri}=0)=0.5$. The censoring time follows an exponential distribution, i.e., $C \sim \text{exp}(0.1)$, which is common in all regions.

\subsection{Estimands} \label{section5.3}
The estimands are the region-specific average RMST differences up to $t^*=4$ concerning the target population. We present its derivation according to Equation~\eqref{eqn5.3} in the Supplementary Materials (Web Appendix E). The average RMST difference is 1.71 in Region 1, 1.51 in Region 2, and 1.15 in Region 3.

\subsection{Methods} \label{section5.4}
First, we consider a Naive estimator using the difference of standard unadjusted RMST between two treatment groups. In Region $r$, the Naive RMST difference is written as follows:
\[
\hat{\Delta}_N(t^*) = \int_0^{t^*} \left\{ \hat{S}_1(t) - \hat{S}_0(t)\right\} dt,
\]
where $\hat{S}_z(t)$ is the standard KM survival curve for Treatment $z$.

Next, we compare the proposed IPSW-adjusted and CW-adjusted estimators for the region-specific average RMST differences. For IPSW-adjusted estimators, with the known distribution of $\bX$ in the target population, we derive the weight $\gamma_{ri}$ as follows:
\[
\gamma_{ri} = \frac{P_{F^*}(\bX = \bX_{ri})}{P_{F_r}(\bX = \bX_{ri})} \propto \frac{P_{F^*}(\bX = \bX_{ri})}{P(d_{ri}=1|X_{ri},R_{ri}=r)P_{F^*}(\bX = \bX_{ri})} = \frac{1}{\rho_r(\bX_{ri})}.
\]
We consider two ways to calculate the sampling scores $\rho_r(\bX) = P(d=1|X,R=r)$. The first uses the true sampling score as known by the data-generating mechanism. The second is estimated by the gradient boosted models~\citep{mccaffrey2013tutorial,burgette2021propensity}. For the CW-adjusted estimators, we make constraints on the first and second moments of $X_1$ and $X_2$ with the constraint function as $\bg(\bX)=[X_1, X_2, X_1^2, X_2^2]$.

Each weighting method is employed to the proposed weighted estimators. For the weighted G-Formula estimator and weighted Augmented estimator, we fit two IPCW RMST regression models in each region:
\begin{eqnarray*}
    \mu_r(t^*|\bX_r,Z_r) &=& \beta_{r0} + \beta_{r1}Z_r + \beta_{r2}X_{r1}  + \beta_{r3}X_{r2} + 
    \beta_{r4}Z_rX_{r1}  + \beta_{r5}Z_rX_{r2}, \\
    \mu_r^{mis}(t^*|\bX_r,Z_r) &=& \beta'_{r0} + \beta'_{r1}Z_r + \beta'_{r2}X_{r1} + \beta'_{r3}Z_rX_{r1}. 
\end{eqnarray*}
Here, we assume the first outcome model is correctly specified while the second is misspecified as it does not include the confounding effect of $X_2$. Accordingly, there are six estimators for each weighting method:
\begin{enumerate}
    \item Weighted Kaplan-Meier estimator (KM).
    \item Weighted G-formula estimator using correctly specified outcome models (GF).
    \item Weighted G-formula estimator using mis-specified outcome models (GFmis).
    \item Weighted Hajek estimator (HJ).
    \item Weighted Augmented estimator using correctly specified outcome models (AG).
    \item Weighted Augmented estimator using mis-specified outcome models (AGmis).
\end{enumerate}
Overall, we have 18 weighted estimators and a Naive estimator.

\subsection{Results} \label{section5.5}
\begin{figure}[!htbp]
\centering\includegraphics[scale = 0.5]{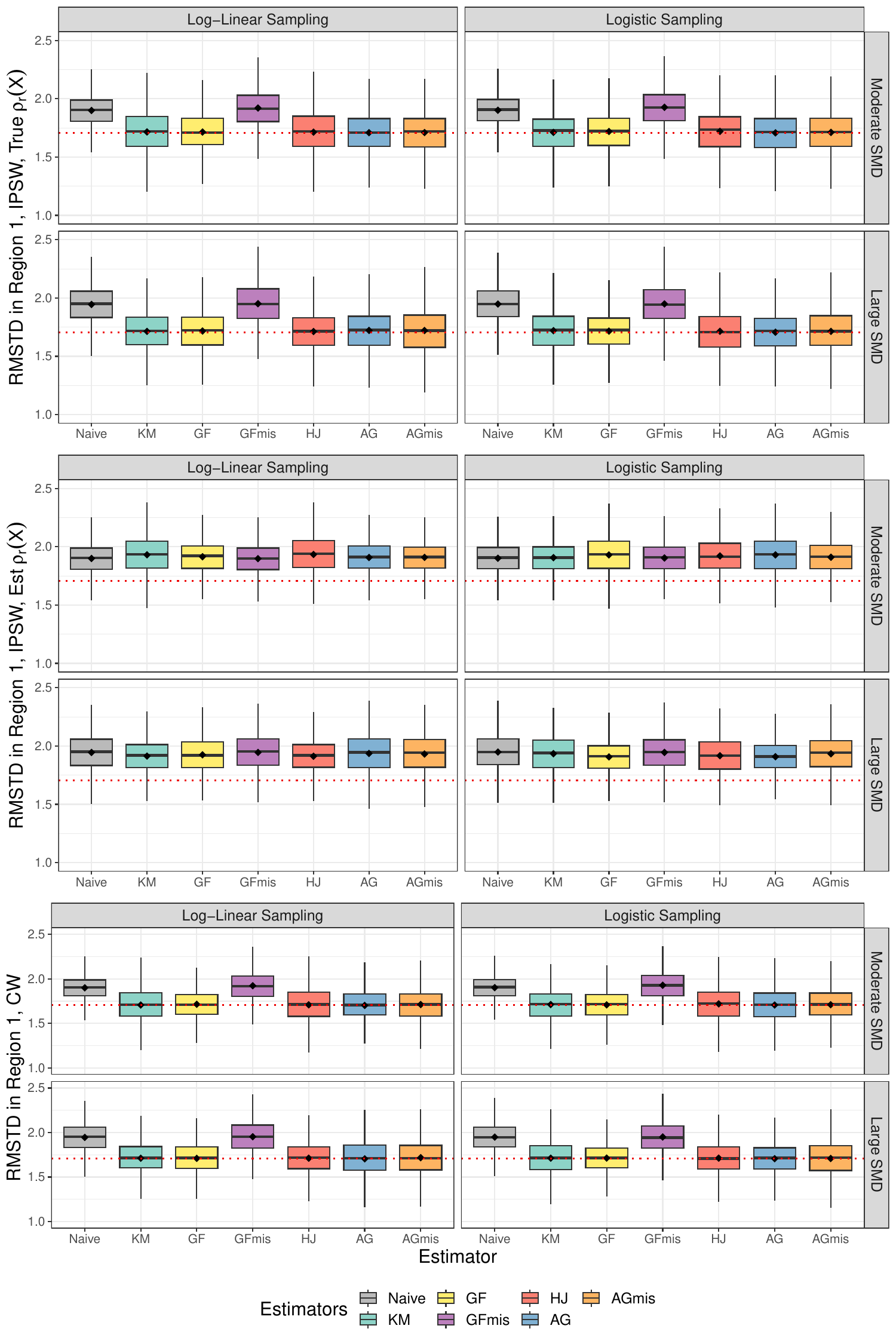}
\caption{Boxplots of estimated average RMST difference (RMSTD) in Region 1 under four sampling scenarios in simulation study. Upper panel: IPSW-adjusted estimators with true sampling score; Middle panel: IPSW-adjusted estimators with estimated sampling score; Bottom panel: CW-adjusted estimators.}
\label{fig2}
\end{figure}

Figure~\ref{fig2} displays the results with box plots of the estimated RMST differences in Region 1 across four scenarios for different sampling score modeling. The first sub-figure shows the results from the IPSW-adjusted estimators under the true sampling score. The second sub-figure shows the results from the IPSW-adjusted estimators under the estimated sampling score. The third sub-figure shows the results from the CW-adjusted estimators. The Naive estimator is included in all panels for reference.

The Naive estimator and all IPSW-adjusted estimators with estimated sampling scores fail to generalize the treatment effects from each region to the target population under all scenarios. These results largely stem from the incorrect estimated sampling score from the gradient boosted model. In contrast, the IPSW-adjusted estimators with the true sampling scores and CW-adjusted estimators yield unbiased estimation across all scenarios, except for the weighted G-Formula estimator when the outcome models are misspecified. The weighted Augmented estimator demonstrates robustness under the outcome model misspecifications. The choice of sampling score models and similarity levels between each region and the target population does not impact bias, but larger SMDs would increase the variance of the estimators. The CW method yields a smaller variance within the same weighted estimator than the IPSW method. Among the five weighted estimators (excluding GFmis), when the outcome models are correctly specified, the weighted G-formula and weighted Augmented estimators exhibit similar and smaller variances than other estimators.

Web Figures 1 and 2 in the Supplementary Materials (Web Appendix E) show the results in Regions 2 and 3 respectively, which present similar patterns as in Region 1.

\vspace*{-2ex}
\section{Case Study} \label{section6}
In this case study, we illustrate our proposed methods using the PLATO dataset under two scenarios: 1) a two-region analysis comparing treatment effects in US vs. non-US regions, and 2) a four-region analysis comparing treatment effects across four predefined geographic regions: i) Asia and Australia, ii) Central and South America, iii) Europe, Middle East, and Africa, and iv) North America. The treatment effect is the RMST difference up to $t^*=360$ days. The results of the four-region analysis are presented in the Supplementary Materials (Web Appendix F).

\subsection{Methods} \label{section6.1}
As we do not have data on the underlying super patient population for PLATO in the entire world, we define the target population as the patient population represented by a simple mixture of the enrolled patients from all regions for the MRCT. As such, we estimate the region-specific average RMST differences by using the proposed weighted estimators concerning the patient population in the pooled dataset. We assume that the treatment-specific propensity score $\pi_{ri}(\bX_{ri})$ is constant based on the allocation rate from the randomized trial in each region. We conduct a univariable effect modifier analysis and select eight binary inessential traits for the weighting methods (detailed in Supplementary Materials, Web Appendix F). The outcome models are fitted by including the selected variables and their interaction with the treatment.

The Naive estimators use the standard unadjusted RMST differences in each region. The IPSW is derived based on Equation~\eqref{eqn3.2} as follows:
\begin{eqnarray*}
    \gamma_{ri} \propto \frac{P(d'_{ri}=1|\bX_{ri})}{P(d_{ri}=1|\bX_{ri},R_{ri}=r)} = \frac{\sum_rP(d_{ri}=1|\bX_{ri},R_{ri}=r)P(R_{ri}=r|\bX_{ri})}{P(d_{ri}=1|\bX_{ri},R_{ri}=r)},
\end{eqnarray*}
where $P(R=r|X)$ is approximated by the proportion of the sample size, $n_r/\sum_rn_r$, for each patient. In the CW-adjusted estimators, we make constraints on the first moment as all covariates are binary.

We use a weighted absolute SMD to evaluate the covariate similarity between each region and the target population before and after applying the weighting methods. Its mathematical definition is provided in the Supplementary Materials (Web Appendix G). Additionally, we conduct a region-specific consistency test of treatment effects for all estimators.

\subsection{Results} \label{section6.2}
\begin{figure}[ht]
\centering\includegraphics[scale = 0.6]{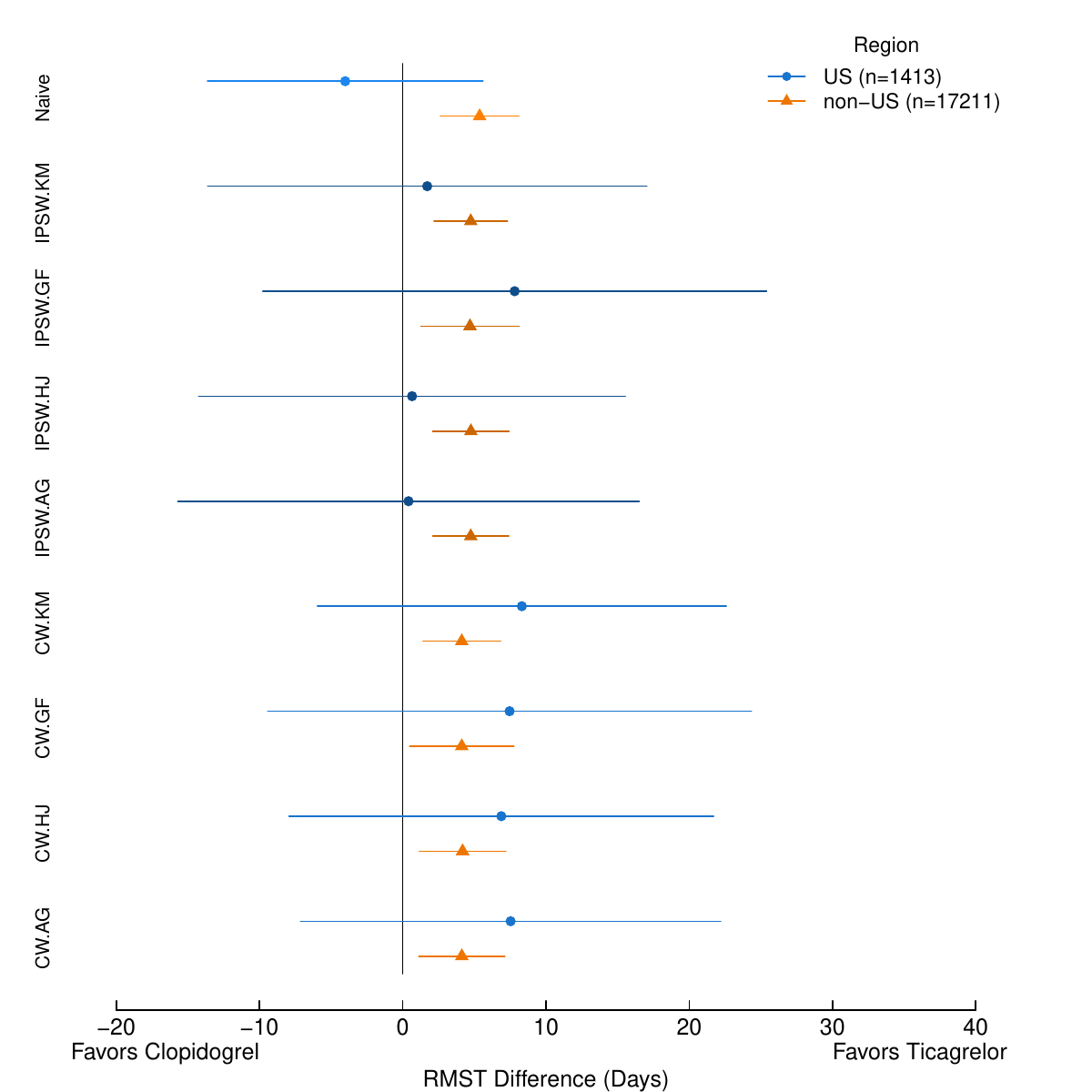}
\caption{Forest plot of estimated average RMST differences with 95\% CIs at $t^*=360$ days from US and non-US in the two-region analysis for PLATO trial.}
\label{fig3}
\end{figure}

Figure~\ref{fig3} and Table~\ref{tab1} present the estimated average RMST differences and the associated 95\% confidence intervals (CI) comparing ticagrelor and clopidogrel for the primary outcome in the two-region analysis. The Naive RMST differences are -4.0 days (95\% CI: -13.6, 5.6) in US and 5.4 days (95\% CI: 2.6, 8.1) in non-US regions, indicating the clopidogrel is more effective than ticagrelor in US, though not statistically significant, while ticagrelor is significantly more effective than clopidogrel in non-US. The consistency test for the Naive estimator shows a strong treatment effect inconsistency between US and non-US, though not significant (p = 0.07).

In the non-US region, the estimated RMST differences are very similar across four estimators within the same weighting method. The IPSW-adjusted estimators yield slightly higher RMST differences than the CW-adjusted estimators (e.g., 4.7 vs. 4.2 days). Both weighting methods indicate that the ticagrelor is significantly more effective than clopidogrel in the non-US region. Under both weighting approaches, the weighted G-formula estimators have the largest variance, potentially due to the misspecification of the fitted RMST regression model. Such misspecification could arise from the RMST regression model not including all confounders or a non-linear association between the selected variables and the outcome.

\begin{table}[ht]
\caption{Estimated average RMST differences with 95\% CIs at $t^*=360$ days from US and non-US in the two-region analysis, and the results from the regional consistency test with p-values and global RMST difference.}
\label{tab1}
\begin{center}
\small
\begin{tabular}{|c|cccc|}
\hline
Method & RMSTD in US & RMSTD in non-US & P-value & Global RMSTD \\
\hline
Naive & -4.0 (-13.6, 5.6) & 5.4 (2.6, 8.1) & 0.07 & 4.7 (2.0, 7.3) \\
IPSW.KM & 2.3 (-12.0, 16.5) & 4.7 (2.2, 7.3) & 0.74 & 4.6 (2.1, 7.2) \\
IPSW.GF & 8.7 (-8.5, 25.9) & 4.7 (1.3, 8.2) & 0.66 & 4.9 (1.5, 8.2) \\
IPSW.HJ & 1.3 (-12.5, 15.0) & 4.7 (2.1, 7.4) & 0.63 & 4.6 (2.0, 7.2) \\
IPSW.AG & 2.2 (-12.5, 17.0) & 4.7 (2.1, 7.4) & 0.74 & 4.7 (2.0, 7.3) \\
CW.KM & 7.4 (-6.0, 20.8) & 4.1 (1.4, 6.9) & 0.64 & 4.3 (1.6, 6.9) \\
CW.GF & 8.3 (-8.3, 24.9) & 4.1 (0.5, 7.8) & 0.63 & 4.3 (0.8, 7.9) \\
CW.HJ & 6.4 (-6.9, 19.7) & 4.2 (1.2, 7.2) & 0.75 & 4.3 (1.4, 7.2) \\
CW.AG & 9.1 (-4.4, 22.5) & 4.2 (1.2, 7.2) & 0.48 & 4.4 (1.5, 7.3) \\
\hline
\end{tabular}
\end{center}
\end{table}

In the US region, the estimated RMST differences vary between the two weighting methods. Among the CW-adjusted estimators, the average RMST differences are close across four estimators, with the highest value in the weighted Augmented estimator at 9.1 days (95\% CI: -4.4, 22.5). These results indicate that ticagrelor is more effective than clopidogrel for the patients in the US after having the selected variable distributions, notably the maintenance aspirin dosage, be resembled in the target population. However, the average RMST differences in the US from the IPSW-adjusted Kaplan-Meier, Hajek, and Augmented estimators are much lower than the corresponding results from the CW-adjusted estimators. For example, the average RMST difference from the IPSW-adjusted Augmented estimator is 2.2 days (95\% CI: -12.5, 17.0) in the US. The IPSW-adjusted G-formula estimator yields similar a result to the CW-adjusted G-formula estimator in the US region.

The consistency tests for all CW-adjusted and IPSW-adjusted estimators reveal no regional treatment effect heterogeneity (see p-values in Table~\ref{tab1}). Consequently, we estimate a global RMST difference by combining the patients from both regions, indicating that ticagrelor is significantly more globally effective than clopidogrel. For example, the global RMST differences from the CW-adjusted and IPSW-adjusted Augmented estimators are 4.4 days (95\% CI: 1.5, 7.3) and 4.7 days (95\% CI: 2.0, 7.3), respectively.

\begin{figure}[!htbp]
\centering\includegraphics[scale = 0.8]{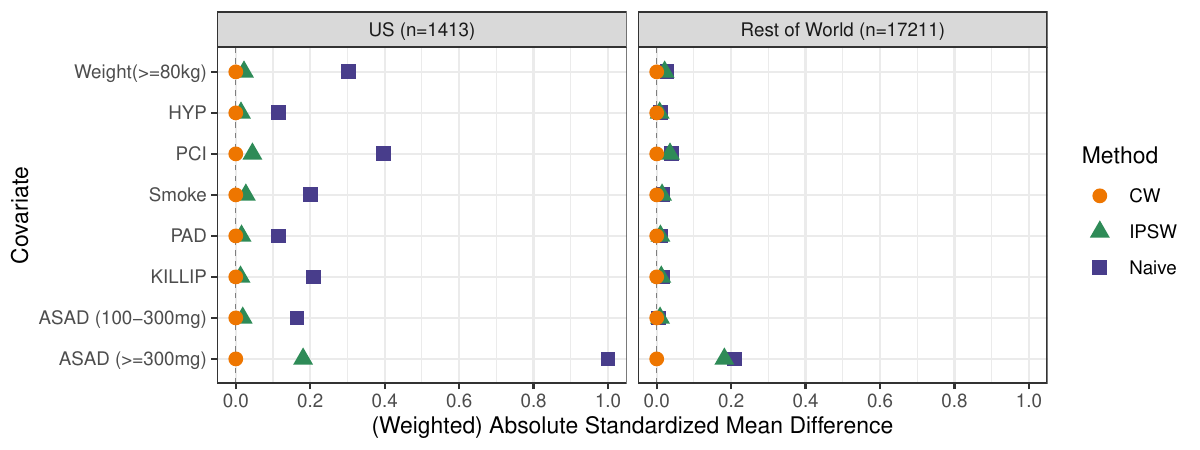}
\caption{Weighted absolute standardized mean differences of eight covariates comparing US and non-US regions to the target population in the two-region analysis for PLATO trial. HYP: hypertension, PCI: percutaneous coronary intervention, PAD: peripheral arterial disease, KILLIP: Killip classification (Level I vs. Level II-IV), ASAD: aspirin dosage.}
\label{fig4}
\end{figure}

In Figure~\ref{fig4}, we present the weighted absolute SMDs of the eight selected covariates in the two-region analysis, comparing US and non-US regions to the target population. Since the non-US population dominates the target population, the absolute SMDs under the Naive estimator between non-US and the target population are close to 0 for all variables, except for the high aspirin dosage group ($\geq 300$ mg). However, the absolute SMDs reveal a notable imbalance of the selected covariates between the US and the target population. In contrast, the absolute SMDs under the CW method are 0 for all variables as the CW method can achieve an exact balance of covariate moments. The IPSW method shows good balances for all variables between the US and the target population except for the high aspirin dosage group ($\geq 300$ mg).

\vspace*{-2ex}
\section{Discussion} \label{section7}
We proposed the calibration weighting (CW) and inverse probability of sampling weighting (IPSW) adjusted estimators to generalize the region-specific treatment effect to a target population in the MRCT. These methods eliminate the disparities arising from the inessential traits. The large sample properties for the CW-adjusted and IPSW-adjusted estimators for the region-specific treatment effects were established. Furthermore, we developed a Wald-type test for the regional consistency test of treatment effect and provided the global treatment effect estimator when the consistency held. Our simulation study demonstrated that the CW-adjusted and IPSW-adjusted estimators with the true sampling scores consistently yielded unbiased estimations across all scenarios. Furthermore, the CW-adjusted estimators exhibited smaller variances compared to the IPSW-adjusted estimators. We highlighted that the IPSW-adjusted estimators are sensitive to the specification of the sampling score models, while the CW-adjusted estimators are more robust. Among the four proposed weighted estimators, the weighted Augmented estimator shows a smaller variance and higher robustness than other estimators.

In the PLATO analysis, our approaches strengthened the previous work that the apparent differences in ticagrelor effects between the US and non-US environment were explained by the difference in the maintenance aspirin dosage and a few other factors. After implementing the CW and IPSW approaches, no significant interaction existed between region and ticagrelor effects. We need to incorporate the difference in these inessential traits to correctly evaluate the region-specific treatment effect, mainly caused by essential traits such as race or genetic variants. We could use other reference populations than the simple mixture of enrolled patients across all regions based on different research questions. For example, if one is interested in knowing how difference the treatment effect of US patients is from non-US patients, we could generalize the treatment effects from US patients to the non-US patient population. Suppose there are no more differences in the treatment effects of these two regions after eliminating the difference from inessential traits. In that case, the US drug regulatory agency may be convinced that the apparent difference before generalization can be attributed to the imbalanced inessential traits (e.g., aspirin dosage) between the two regions. If the region difference still exists, it may be attributed to other hidden region-specific factors (essential traits), such as racial and genetic differences between the two regions.

There are several avenues for future research to enhance our proposed method. First, when unobserved confounders have associations with both treatment effect and sampling scores, further research is required to reduce the bias caused by such factors. Second, we selected the effect modifiers in the PLATO analysis by fitting univariable Cox regression and RMST regression models. A more robust criteria model needs to be further investigated for high dimensional covariates to refine our approach. Third, our proposed methods could be extended to the MRCT design by incorporating power analysis, sample size calculation, and type I error control.

Our proposed method allows rigorous assessment of region-specific treatment effects against the target population and the treatment effect heterogeneity across regions by eliminating the potential effect modifications of imbalanced baseline characteristics across regions in an MRCT. While our discussion of this work is in the context of MRCT analysis, it is worthy to note that this method can be applied to assessing treatment effect heterogeneity across patient subgroups for which the effect modifying covariates are not balanced across subgroups in randomized clinical trials.

\vspace*{-2ex}
\section*{Software}
The relevant R code for the methodology and simulation study is available on \url{https://github.com/kimihua1995/CW_MRCT_RMST}.

\vspace*{-2ex}
\section*{Data Availability Statement}
The PLATO data is available to members of the PLATO executive committee. Without the permission of the third parties and to avoid unintended leakage of patient privacy, we elect to not share the data. Individual investigators may reach out directly to the PLATO executive committee for collaboration.

\vspace*{-2ex}
\section*{Acknowledgement}
Dr. Hong was partially supported by the National Institute of Mental Health (R01 MH126856) and Patient-Centered Outcomes Research Institute (ME-2020C3-21145). Dr. Wang was partially supported by the NCI (P01 CA142538) and the NIA (R01 AG066883). The PLATO trial was supported by AstraZeneca. We appreciate the statistical and clinical insights of Lars Elvelin, PhD, Wilhelm Ridderstrale, PhD, and Mikael Knutsson, PhD from AstraZeneca.

\bibliographystyle{agsm.bst}
\bibliography{Reference.bib}

\end{document}


\begin{center}
{\large{ \textbf{Supplementary Materials for ``Inference of treatment effect and its regional modifiers using restricted mean survival time in mutli-regional clinical trials"}}} \medskip

\end{center}

\bigskip

The Supplementary Material is organized as follows. Web Appendix A shows the derivation of the calibration weight $\hat{p}_{ri}$ from the convex optimization problem using the Lagrange multiplier. In Web Appendix B to D, we provide the proofs of Theorems 1 to 4 for the proposed weighted estimators of region-specific average RMST difference. Web Appendix E includes the additional simulation set-ups and results. We show the variable selection and results of the four-region analysis for the PLATO trial from the case study in Web Appendix F. Web Appendix G introduces the definition of the weighted absolute standardized mean difference.

\section*{Web Appendix A: Derivation of Calibration Weights}
\noindent The objective function from the convex problem as discussed in Section 3.2 is
\[
L(\blambda) = \sum_{r=1}^{M}\sum_{i=1}^{n_r}p_{ri}log(p_{ri}) - \sum_{r=1}^{M}\blambda_r^T \left \{ \sum_{i=1}^{n_r}p_{ri}\bg(\bX_{ri}) - \Tilde{\bg} \right \} - \sum_{r=1}^{M}\lambda_{r0} \left \{ \sum_{i=1}^{n_r}p_{ri} - 1 \right \}.
\]
To minimize $L(\blambda)$, for $i = 1,\dots,n_r$, let $\frac{\partial L}{\partial p_{ri}} = 1 + log(p_{ri}) - \blambda_r^T \bg(\bX_{ri}) - \lambda_{r0} = 0$, we get $p_{ri} = \text{exp}\{\blambda_r^T \bg(\bX_{ri})\}/\text{exp}\{1-\lambda_{r0}\}$. Since $\sum_{i=1}^{n_r}p_{ri} = 1$, then $\text{exp}\{1-\lambda_{r0}\} = \sum_{i=1}^{n_r}\text{exp}\{\blambda_r^T\bg(\bX_{ri})\}$. Therefore, 
\[
\hat{p}_{ri} = \frac{\text{\text{exp}}\{\blambda_r^T\bg(\bX_{ri})\}}{\sum_{i=1}^{n_r}\text{exp}\{\blambda_r^T\bg(\bX_{ri})\}}.
\]
Since $\sum_{i=1}^{n_r}p_{ri}\bg(\bX_{ri}) = \Tilde{\bg}$, after plugging in the $\hat{p}_{ri}$, we have
\[
\sum_{i=1}^{n_r}\text{exp}\{\blambda_r^T\bg(\bX_{ri})\}\{\bg(\bX_{ri}) - \Tilde{\bg}\} = 0.
\]

\section*{Web Appendix B: Proofs of Theorem 1}
\noindent We will first show the proof for Theorem 1 concerning the calibration weights $\hat{p}_{ri}$ (i.e., the CW-adjusted Kaplan-Meier estimator). The proofs for the inverse probability of sampling weights, $\hat{\gamma}_{ri}$, are similar, and we will provide some discussions on the key steps.

\subsection*{B.1 Notations}
\noindent For $i = 1,\dots,n_{r}$, we let the calibration weighted counting process and at risk process for Treatment $z$ in Region $r$ be
\vspace*{-2ex}
\begin{eqnarray*}
    \Tilde{N}_{rz}(t) &=& \sum_{i=1}^{n_{r}}\hat{p}_{ri}\hat{q}_{rzi}N_{rzi}(t), \\
    \Tilde{Y}_{rz}(t) &=& \sum_{i=1}^{n_{r}}\hat{p}_{ri}\hat{q}_{rzi}Y_{rzi}(t),
\end{eqnarray*}
where $N_{rzi}(t) = I[Y_{ri} \leq t; \delta_{ri}=1; Z_{ri}=z]$ and $Y_{rzi}(t) = I[Y_{ri} \geq t; Z_{ri}=z]$. We denote $\hat{q}_{rzi} = \frac{I(Z_{ri} = z)}{\hat{\pi}_{ri}(\bX_{ri})^z(1 - \hat{\pi}_{ri}(\bX_{ri}))^{1-z}}$ as the estimated inverse individual treatment-specific propensity score for $q_{rzi}(\bX_{ri})$. Suppose the process $M_{rzi}(t) = N_{rzi}(t) - \int_0^tY_{rzi}(u)d\Lambda_{rz}(u)$, where $\Lambda_{rz}(t) = E_{\bX}[-\text{log} \{ S(t|\bX, Z = z, R = r, d = 1) \} ]$ is the cumulative incidence function. $M_{rzi}(t)$ is a martingale with its derivation $dM_{rzi}(t) = dN_{rzi}(t) - Y_{rzi}(u)d\Lambda_{rz}(u)$. We define the entropy weighted process $\Tilde{M}_{rz}(t) = \Tilde{N}_{rz}(t) - \int_0^t\Tilde{Y}_{rz}(u)d\Lambda_{rz}(u)$, with its derivation:
\vspace*{-2ex}
\begin{eqnarray*}
    d\Tilde{M}_{rz}(t) &=& d\Tilde{N}_{rz}(t) - \Tilde{Y}_{rz}(t)d\Lambda_{rz}(t) \\
    &=& \sum_{i=1}^{n_{r}}\hat{p}_{ri}\hat{q}_{rzi}[dN_{rzi}(t) - Y_{rzi}(t)d\Lambda_{rz}(t)] \\
    &=& \sum_{i=1}^{n_{r}}\hat{p}_{ri}\hat{q}_{rzi}dM_{rzi}(t).
\end{eqnarray*}
For $r = 1,\dots,M$ and $z = 0,1$, let $\omega_{rz}(t) = P(Y_{ri} \geq t | Z_{ri} = z)$ and $\bar{Y}_{rz}(t) = \sum_{i=1}^{n_r}\hat{q}_{rzi}Y_{rzi}(t)$. We assume that for any fixed $0 < t^* < \infty$, as $n_r \to \infty$, \\ $\sup\limits_{0 \leq t \leq t^*}|\frac{\bar{Y}_{rz}(t)}{\sum_{i=1}^{n_r}I[Z_{ri}=z]} - \omega_{rz}(t)| \overset{p}{\to} 0$. By $\overset{p}{\to}$ we mean ``converges in probability".
\vspace*{10pt}

\subsection*{B.2 Lemma 1}
\noindent \textbf{Lemma 1.} \textit{Assume the sampling score of trial participation in each region is proportional to a log-linear model with respect to $\bg(\bX)$, that is, $\rho_r(\bX) \propto \text{exp}\{\bet_r^T\bold{g}(\bX)\}$, the estimated calibration weights satisfy $\hat{p}_{ri} - (N_r \rho_r(\bX_i))^{-1} \overset{p}{\to} 0$, as $n_r \to \infty$.}

\textbf{Proof}. Let $d_{ri} = 1$ for trial participants and $d_{ri} = 0$ for non-participants in region $r$, then $E[d_{r}/\rho_r(\bX)] = 1$. Let $\bmu_{g0} = E[\bg(\bX)]$, by using the M-estimator theory, we write the objective function, $L(\blambda)$, proposed in Web Appendix A as the following estimating equations:
\begin{eqnarray}
\frac{1}{N_r}\sum_{i=1}^{N_r}\xi_1(\bX_i,d_{ri};\bmu_g) &=& \frac{1}{N_r}\sum_{i=1}^{N_r}d_{ri}\{\bg(\bX_i) - \bmu_g\} = 0. \\
\frac{1}{N_r}\sum_{i=1}^{N_r}\xi_2(\bX_i,d_{ri};\bmu_g) &=& \frac{1}{N_r}\sum_{i=1}^{N_r}d_{ri}\text{exp}\{\blambda_r^T\bg(\bX_i)\}\{\bg(\bX_i) - \bmu_g\} = 0.
\end{eqnarray}
Notice that $\bmu_{g0}$ is the solution to $E[\xi_1(\bX,d_{r};\bmu_g)] = 0$. Taking expectation on the left hand side of Equation A.2 with $\bmu_g = \bmu_{g0}$ and under the assumption of \\ $\rho_r(\bX) \propto \text{exp}\{\bet_r^T\bold{g}(\bX)\}$ leads to
\vspace*{-2ex}
\begin{eqnarray*}
E[\xi_2(\bX,d_{r};\bmu_{g0})] &=& E \left [d_{r}\text{exp}\{\blambda_r^T\bg(\bX)\}\{\bg(\bX) - \bmu_{g0}\} \right ] \\
&=& E \left [ E[d_{r}\text{exp}\{\blambda_r^T\bg(\bX)\}\{\bg(\bX) - \bmu_{g0}\} | \bX] \right ] \\
&=& E \left [ \text{exp}\{\blambda_r^T\bg(\bX)\}\{\bg(\bX) - \bmu_{g0}\} Pr(A=1|\bX, R=r) \right ] \\
&\propto& E \left [ \text{exp}\{(\blambda_r + \bet_r)^T\bg(\bX)\}\{\bg(\bX) - E[\bg(\bX)]\} \right ].
\end{eqnarray*}
For the above expectation to be zero, one needs $\text{exp}\{(\blambda_r + \bet_r)^T\bg(\bX)\}$ to be a constant, that is, $\bet_r = -\blambda_r$. Therefore,
\vspace*{-2ex}
\begin{eqnarray*}
\hat{p}_{ri}[N_r \rho_r(\bX_{ri})] &=& \frac{\text{exp}\{\blambda_r^T\bg(\bX_{ri})\}}{\sum_{i=1}^{n_r}\text{exp}\{\blambda_r^T\bg(\bX_{ri})\}} [N_r \rho_r(\bX_{ri})] \\
&=& \frac{N_r}{\sum_{i=1}^{N_r}\text{exp}\{\blambda_r^T\bg(\bX_{ri})\}d_{ri}}\\
&=& \left. N_r \middle/ \sum_{i=1}^{N_r}\frac{d_{ri}}{\rho_r(\bX_{ri})} \right. \\
&\overset{p}{\to}& 1.
\end{eqnarray*}

\noindent In Lemma 1, we made a log-linear assumption on the sampling score to show the validity of the CW-adjusted estimators in the following materials. Under this assumption, we showed that there is a direct correspondence between the calibration weights $\hat{p}_{ri}$ and the estimated sampling score, that is, $\hat{p}_{ri} - (N_r \rho_r(\bX_i))^{-1} \overset{p}{\to} 0$. The model assumption of the sampling score is related to the objective function we use in the optimization problem, i.e., the entropy function. We can also assume that the sampling score follows a logistic regression model with $\sum_{r=1}^{M}\sum_{i=1}^{n_r}(p_{ri} - 1)\text{log}(p_{ri} - 1)$ as the objective function (Josey et al., 2021). And it can be shown that the log-linear model is close to the logistic model when the fraction of trial participants to the reference population, i.e., $n_r/N_r$, is small (Lee et al., 2023). Note that the log-linear sampling score assumption simplifies the proofs of the large sample properties of the CW-adjusted estimators, but calculating the calibration weights does not require the estimation of the unknown sampling scores. To validate its robustness, we showed in the simulation study that the CW-adjusted estimators were unbiased under both the log-linear and the logistic sampling score model. A less strict assumption remains in future research.
\vspace*{10pt}

\subsection*{B.3 Large Sample Properties of CW-Adjusted Survival Function}
\noindent Before we show the large sample properties of the CW-adjusted KM estimator of the region-specific average RMST, we first show the large sample properties of the CW-adjusted KM estimator of the region-specific average survival function, $\Tilde{S}_{rz}$, where
\begin{eqnarray*}
    \Tilde{S}_{rz}(t) = \prod_{u \leq t} \left \{ 1 - \frac{d\Tilde{N}_{rz}(u)}{\Tilde{Y}_{rz}(u)} \right \}
\end{eqnarray*}
Let $S_{rz}(t) = E_{F^*}[S_{rz}(t|\bX)]$ denote the region-specific average survival function, we have the following theorem for $\Tilde{S}_{rz}$: \\
\textbf{Theorem 1*}. \textit{For any fixed $0 < t^* < \infty$, as $n_r \to \infty$,
\vspace*{-2ex}
\begin{eqnarray*}
    \sup\limits_{0 \leq t \leq t^*}\left |\Tilde{S}_{rz}(t) - S_{rz}(t) \right| &\overset{p}{\to}& 0, \\
    \sqrt{n_r}\left\{ \Tilde{S}_{rz}(t) - S_{rz}(t) \right\} &\overset{d}{\to}& N \left( 0, S_{rz}^2(t)\sigma_{rz}^2(t) \right),
\end{eqnarray*}
where $\sigma_{rz}^2(t) = n_{rz}\int_0^t\frac{d\Lambda_{rz}(u)}{\Tilde{W}_{rz}(u)}$ and $\Lambda_{rz}(t) = E_{\bX}[-\text{log} \{ S(t|\bX, Z = z, R = r, d = 1) \} ]$ is the cumulative incidence function. Let $\hat{\sigma}^2_{rz}(t) = \int_0^t\frac{n_rd\Tilde{N}_{rz}(u)}{\Tilde{W}_{rz}(u)(\Tilde{Y}_{rz}(u) - \Delta \Tilde{N}_{rz}(u))}$, as $n_r \to \infty$,
\vspace*{-2ex}
\begin{eqnarray*}
\sup\limits_{0 \leq t \leq t^*}|\hat{\sigma}^2_{rz}(t) - \sigma_{rz}^2(t)| \overset{p}{\to} 0.
\end{eqnarray*}
}

\noindent \textbf{\textit{Proof of the consistency in Theorem 1*.}} \\
If $S_{rz}(t) > 0$, by the formulas for integration by parts (referred by the proof of Theorem 3.2.3 in Fleming and Harrington, 1991)
\vspace*{-2ex}
\begin{eqnarray*}
\frac{\Tilde{S}_{rz}(t)}{S_{rz}(t)} &=& 1 - \int_0^t\frac{\Tilde{S}_{rz}(u-)}{S_{rz}(u)}\left \{ \frac{d\Tilde{N}_{rz}(u)}{\Tilde{Y}_{rz}(u)} - d\Lambda_{rz}(u) \right \}. \\
\frac{S_{rz}(t) - \Tilde{S}_{rz}(t)}{S_{rz}(t)} &=& \int_0^t\frac{\Tilde{S}_{rz}(u-)}{S_{rz}(u)}\left \{ \frac{d\Tilde{N}_{rz}(u)}{\Tilde{Y}_{rz}(u)} - d\Lambda_{rz}(u) \right \} \\
&=& \int_0^t\frac{\Tilde{S}_{rz}(u-)}{S_{rz}(u)}\left \{ \frac{d\Tilde{N}_{rz}(u)}{\Tilde{Y}_{rz}(u)} - \frac{\Tilde{Y}_{rz}(u)}{\Tilde{Y}_{rz}(u)}I[\Tilde{Y}_{rz}(u) > 0]d\Lambda_{rz}(u) \right. \\
&-& \left. I[\Tilde{Y}_{rz}(u) = 0]d\Lambda_{rz}(u)\right \}\\
&=& \int_0^t\frac{\Tilde{S}_{rz}(u-)}{S_{rz}(u)}\frac{I[\Tilde{Y}_{rz}(u) > 0]}{\Tilde{Y}_{rz}(u)}d\Tilde{M}_{rz}(u) - B_{rz}(t),
\end{eqnarray*}
where $B_{rz}(t) = \int_0^t\frac{\Tilde{S}_{rz}(u-)}{S_{rz}(u)}I[\Tilde{Y}_{rz}(u) = 0]d\Lambda_{rz}(u)$. Let $\tau = \text{inf}\{u: \Tilde{Y}_{rz}(u) = 0\}$ and $\Tilde{B}_{rz}(t) = B_{rz}(t)S_{rz}(t)$. Then $\forall u > \tau$, $\Tilde{Y}_{rz}(u) = 0$,
\vspace*{-2ex}
\begin{eqnarray*}
\Tilde{B}_{rz}(t) &=& S_{rz}(t)I[\tau < t]\int_\tau^t\frac{\Tilde{S}_{rz}(u-)}{S_{rz}(u)}d\Lambda_{rz}(u) \\
&=& S_{rz}(t)I[\tau < t]\int_\tau^t\frac{\Tilde{S}_{rz}(u-)}{S_{rz}(u)}\frac{-dS_{rz}(u)}{S_{rz}(u-)} \\
&=& S_{rz}(t)I[\tau < t]\Tilde{S}_{rz}(\tau)\int_\tau^td\frac{1}{S_{rz}(u)} \\
&=& I[\tau < t]\frac{\Tilde{S}_{rz}(\tau)\{S_{rz}(\tau) - S_{rz}(t)\}}{S_{rz}(\tau)}.
\end{eqnarray*}
As $n_{r} \to \infty$,
\vspace*{-2ex}
\begin{eqnarray*}
E[\Tilde{B}_{rz}(t)] &\leq& E\left[I[\tau < t]\{1 - \frac{S_{rz}(t)}{S_{rz}(\tau)}\} \right] \\
&\leq& E\left[I[\tau < t]\{1 - S_{rz}(t)\} \right] \\
&=& \{1 - S_{rz}(t)\}P(\Tilde{Y}_{rz}(t) = 0) \\
&=& \{1 - S_{rz}(t)\}\{1 - \omega_{rz}(t) \}^{\sum_{i=1}^{n_r}I[Z_{ri}=z]} \\
&\to& 0.
\end{eqnarray*}
This implies that as $n_{r} \to \infty$, $B_{rz}(t) \overset{p}{\to} 0$. Therefore, for any fixed $0 < t^* < \infty$,
\begin{eqnarray*}
P \left( \frac{S_{rz}(t) - \Tilde{S}_{rz}(t)}{S_{rz}(t)} = V_{rz}(t), t \in [0,t^*] \right) \to 1,
\end{eqnarray*}
where $V_{rz}(t) = \int_0^t\frac{\Tilde{S}_{rz}(u-)}{S_{rz}(u)}\frac{I[\Tilde{Y}_{rz}(u) > 0]}{\Tilde{Y}_{rz}(u)}d\Tilde{M}_{rz}(u)$. 

\noindent Based on Lemma 1,
\vspace*{-2ex}
\begin{eqnarray*}
V_{rz}(t) &=& \sum_{i=1}^{N_r}\int_0^t\frac{\Tilde{S}_{rz}(u-)}{S_{rz}(u)}\frac{I[\Tilde{Y}_{rz}(u) > 0]}{\Tilde{Y}_{rz}(u)}d_{ri}\hat{p}_{ri}\hat{q}_{rzi}dM_{rzi}(u) \\
&\overset{p}{\to}& \frac{1}{N_r}\sum_{i=1}^{N_r}\int_0^t\frac{\Tilde{S}_{rz}(u-)}{S_{rz}(u)}\frac{I[\Tilde{Y}_{rz}(u) > 0]}{\Tilde{Y}_{rz}(u)}\frac{d_{ri}}{\rho_r(\bX_{ri})}\hat{q}_{rzi}dM_{rzi}(u) \\
&\overset{p}{\to}& \frac{1}{N_r}\sum_{i=1}^{N_r}\int_0^t\frac{\Tilde{S}_{rz}(u-)}{S_{rz}(u)}\frac{I[\bar{Y}_{rz}(u) > 0]}{\omega_{rz}(u)}\hat{q}_{rzi}\frac{d_{ri}}{\rho_r(\bX_{ri})}dM_{rzi}(u) \\
&\overset{p}{\to}& E\left[ \int_0^t\frac{\Tilde{S}_{rz}(u-)}{S_{rz}(u)}\frac{I[\bar{Y}_{rz}(u) > 0]}{\omega_{rz}(u)}dM_{rzi}(u)\right],
\end{eqnarray*}
where $\bar{Y}_{rz}(u) = \sum_{i=1}^{n_r}\hat{q}_{rzi}Y_{rzi}(u)$. The second ``$\overset{p}{\to}$" is derived as the following:
\begin{enumerate}
    \item Since $\hat{p}_{ri} >0$, $I[\Tilde{Y}_{rz}(u) > 0] = I[\sum_{i=1}^{n_r}\hat{p}_{ri}\hat{q}_{rzi}Y_{rzi}(u) > 0] = $ \\
    $I[\sum_{i=1}^{n_r}\hat{q}_{rzi}Y_{rzi}(u) > 0] = I[\bar{Y}_{rz}(u) > 0]$.
    \item $\Tilde{Y}_{rz}(u) = \sum_{i=1}^{N_r}d_{ri}\hat{p}_{ri}\hat{q}_{rzi}I[Y_{rzi} \geq u] \overset{p}{\to} \frac{1}{N_r}\sum_{i=1}^{N_r}\frac{d_{ri}}{\rho_r(X_i)}\hat{q}_{rzi}I[Y_{rzi} \geq u] \overset{p}{\to} P(Y_{rzi} \geq u) = \omega_{rz}(u)$.
\end{enumerate}
We let
\vspace*{-2ex}
\begin{eqnarray*}
V^*_{rz}(t) &=& \frac{1}{n_r}\sum_{i=1}^{n_r}\int_0^t\frac{\Tilde{S}_{rz}(u-)}{S_{rz}(u)}\frac{I[\bar{Y}_{rz}(u) > 0]}{\omega_{rz}(u)}\hat{q}_{rzi}dM_{rzi}(u) \\
&\overset{p}{\to}& E\left[ \int_0^t\frac{\Tilde{S}_{rz}(u-)}{S_{rz}(u)}\frac{I[\bar{Y}_{rz}(u) > 0]}{\omega_{rz}(u)}dM_{rzi}(u)\right],
\end{eqnarray*}
then $\sup\limits_{0 \leq t \leq t^*}|V_{rz}(t) - V^*_{rz}(t)| \overset{p}{\to} 0$. Therefore, to show consistency of $\Tilde{S}_{tz}(t)$, it is sufficient to show $\sup\limits_{0 \leq t \leq t^*}|V^*_{rz}(t)|^2 \overset{p}{\to} 0$. Since the process $M_{rzi}(u)$ is a martingale and the process $\frac{\Tilde{S}_{rz}(u-)}{S_{rz}(u)}\frac{I[\bar{Y}_{rz}(u) > 0]}{\omega_{rz}(u)}\hat{q}_{rzi}$ is predictable and bounded, then by Corollary 3.4.1 in Fleming and Harrington (1991) and the assumption of $\sup\limits_{0 \leq t \leq t^*}|\frac{\bar{Y}_{rz}(t)}{\sum_{i=1}^{n_r}I[Z_{ri}=z]} - \omega_{rz}(t)| \overset{p}{\to} 0$, $\forall \epsilon, \eta > 0$,
\vspace*{-2ex}
\begin{eqnarray*}
&& P\{\sup\limits_{0 \leq t \leq t^*}|V^*_{rz}(t)|^2 \geq \epsilon\} \\
&\leq& \frac{\eta}{\epsilon} + P\{\sum_{i=1}^{n_r}\int_0^{t^*}\frac{\Tilde{S}^2_{rz}(u-)}{S^2_{rz}(u)}\frac{I[\bar{Y}_{rz}(u) > 0]}{n_r^2\omega^2_{rz}(u)}\hat{q}_{rzi}^2Y_{rzi}(u)d\Lambda_{rz}(u) \geq \eta\} \\
&\leq& \frac{\eta}{\epsilon} + P\{\int_0^{t^*}\frac{\Tilde{S}^2_{rz}(u-)}{S^2_{rz}(u)}\frac{I[\bar{Y}_{rz}(u) > 0]}{n_r^2\omega^2_{rz}(u)}\bar{Y}_{rz}(u)d\Lambda_{rz}(u) \geq \eta\} \\
&\leq& \frac{\eta}{\epsilon} + P\{\frac{\Lambda_{rz}(t^*)\sum_{i=1}^{n_r}I[Z_{ri}=z]}{S^2_{rz}(t^*)n_r^2\omega_{rz}(u)} \geq \eta \}.
\end{eqnarray*}
By letting $\eta \to 0$, the right-hand side of the above converge to 0 as $n_{r} \to \infty$. Therefore, as $n_{r} \to \infty$, $\sup\limits_{0 \leq t \leq t^*}|V_{rz}(t)| \overset{p}{\to} 0$, thus $\sup\limits_{0 \leq t \leq t^*}|\Tilde{S}_{rz}(t) - S_{rz}(t)| \overset{p}{\to} 0$.
\vspace*{10pt}

\noindent \textbf{\textit{Proof of the asymptotic normality in Theorem 1*.}} We at first show a corollary.

\noindent \textbf{Corollary 1.} \textit{Let $\Tilde{\Lambda}_{rz}(t) = \int_0^t\frac{d\Tilde{N}_{rz}(u)}{\Tilde{Y}_{rz}(u)}$, for any fixed $0 < t^* < \infty$, as $n_r \to \infty$,
\vspace*{-2ex}
\begin{eqnarray*}
\sup\limits_{0 \leq t \leq t^*}|\Tilde{\Lambda}_{rz}(t) - \Lambda_{rz}(t)| \overset{p}{\to} 0.
\end{eqnarray*}
}

\textbf{Proof.} For any $t \in [0,t^*]$,
\vspace*{-2ex}
\begin{eqnarray*}
\left| \Tilde{\Lambda}_{rz}(t) - \Lambda_{rz}(t) \right| &=& \left| \int_0^t\frac{d\Tilde{N}_{rz}(u)}{\Tilde{Y}_{rz}(u)} - \int_0^td\Lambda_{rz}(u) \right| \\
&\leq& \left| \int_0^t\frac{d\Tilde{N}_{rz}(u)}{\Tilde{Y}_{rz}(u)} - \int_0^t\frac{\Tilde{Y}_{rz}(u)}{\Tilde{Y}_{rz}(u)}I[\Tilde{Y}_{rz}(u) > 0]d\Lambda_{rz}(u) \right| \\
&+& \left| \int_0^tI[\Tilde{Y}_{rz}(u) = 0]d\Lambda_{rz}(u)  \right| \\
&\leq&  \left|\int_0^t\frac{I[\Tilde{Y}_{rz}(u) > 0]}{\Tilde{Y}_{rz}(u)}d\Tilde{M}_{rz}(u) \right| + I[\Tilde{Y}_{rz}(t) = 0]\Lambda_{rz}(t).
\end{eqnarray*}
Note that $I[\Tilde{Y}_{rz}(t) = 0]\Lambda_{rz}(t) \overset{p}{\to} 0$ as $n_r \to \infty$. We let $Q_{rz}(t) = \int_0^t\frac{I[\Tilde{Y}_{rz}(u) > 0]}{\Tilde{Y}_{rz}(u)}d\Tilde{M}_{rz}(u)$.
\vspace*{-2ex}
\begin{eqnarray*}
Q_{rz}(t) &=& \sum_{i=1}^{N_r}\int_0^t\frac{I[\Tilde{Y}_{rz}(u) > 0]}{\Tilde{Y}_{rz}(u)}d_{ri}\hat{p}_{ri}\hat{q}_{rzi}dM_{rzi}(u) \\
&\overset{p}{\to}& \frac{1}{N_r}\sum_{i=1}^{N_r}\int_0^t\frac{I[\Tilde{Y}_{rz}(u) > 0]}{\Tilde{Y}_{rz}(u)}\frac{d_{ri}}{\rho(\bX_i)}d_{ri}\hat{p}_{ri}\hat{q}_{rzi}dM_{rzi}(u) \\
&\overset{p}{\to}& E\left[ \int_0^t\frac{I[\bar{Y}_{rz}(u) > 0]}{\omega_{rz}(u)}dM_{rzi}(u)\right].
\end{eqnarray*}
We let $Q_{rz}^*(t) = \frac{1}{n_r}\sum_{i=1}^{n_r}\int_0^t\frac{I[\bar{Y}_{rz}(u) > 0]}{\omega_{rz}(u)}\hat{q}_{rzi}dM_{rzi}(u)$, then $\sup\limits_{0 \leq t \leq t^*}|Q_{rz}(t) - Q_{rz}^*(t)| \overset{p}{\to} 0$. By Corollary 3.4.1 in Fleming and Harrington (1991), $\forall \epsilon, \eta > 0$,
\vspace*{-2ex}
\begin{eqnarray*}
P\{\sup\limits_{0 \leq t \leq t^*}|Q^*_{rz}(t)|^2 \geq \epsilon\} &\leq& \frac{\eta}{\epsilon} + P\{\int_0^{t^*}\frac{I[\bar{Y}_{rz}(u) > 0]}{n_r^2\omega^2_{rz}(u)}\bar{Y}_{rz}(u)d\Lambda_{rz}(u) \geq \eta\} \\
&\leq& \frac{\eta}{\epsilon} + P\{\frac{\Lambda_{rz}(t^*)\sum_{i=1}^{n_r}I[Z_{ri}=z]}{n_r^2\omega_{rz}(u)} \geq \eta \}.
\end{eqnarray*}
By letting $\eta \to 0$, the right-hand side of the above converges to 0 as $n_r \to \infty$. Therefore, as $n_r \to \infty$, $\sup\limits_{0 \leq t \leq t^*}|Q_{rz}(t)| \overset{p}{\to} 0$, thus $\sup\limits_{0 \leq t \leq t^*}|\Tilde{\Lambda}_{rz}(t) - \Lambda_{rz}(t)| \overset{p}{\to} 0$.
\vspace*{10pt}

\noindent Now back to the proof of the asymptotic normality, as shown above,
\begin{eqnarray*}
\sqrt{n_r}\left\{ S_{rz}(t) - \Tilde{S}_{rz}(t) \right\} &\overset{p}{\to}& S_{rz}(t)\sqrt{n_r}\sum_{i=1}^{N_r}\int_0^t\frac{\Tilde{S}_{rz}(u-)}{S_{rz}(u)}\frac{I[\Tilde{Y}_{rz}(u) > 0]}{\Tilde{Y}_{rz}(u)}d_{ri}\hat{p}_{ri}\hat{q}_{rzi}dM_{rzi}(u).
\end{eqnarray*}
We define $H_{rzi}(t) = \sqrt{n_r}\frac{\Tilde{S}_{rz}(u-)}{S_{rz}(u)}\frac{I[\Tilde{Y}_{rz}(u) > 0]}{\Tilde{Y}_{rz}(u)}d_{ri}\hat{p}_{ri}\hat{q}_{rzi}$, $U_{rz}(t) = \sum_{i=1}^{N_r}\int_0^tH_{rzi}(u)dM_{rzi}(u)$ and $U_{rz,\epsilon}(t) = \sum_{i=1}^{N_r}\int_0^tH_{rzi}(u)I[|H_{rzi}(u)| \geq \epsilon]dM_{rzi}(u)$. Let $<\cdot,\cdot>$ denote predictable variation process of martingale, then as $n_r \to \infty$,
\begin{eqnarray*}
<U_{rz}, U_{rz}>(t) &=& n_r\sum_{i=1}^{N_r}\int_0^tH_{rzi}^2(u)Y_{rzi}(u)d\Lambda_{rz}(u) \\
&=& n_r\sum_{i=1}^{n_r}\int_0^t\frac{\Tilde{S}_{rz}^2(u-)}{S_{rz}^2(u)}\frac{I[\Tilde{Y}_{rz}(u) > 0]}{\Tilde{Y}_{rz}^2(u)}\hat{p}^2_{ri}[\hat{q}_{rzi}]^2Y_{rzi}(u)d\Lambda_{rz}(u) \\
&\overset{p}{\to}& n_r\int_0^{t}\frac{\sum_{i=1}^{n_r}[\hat{p}_{ri}I[Z_{ri}=z]/\pi_{Zi}^z(1 - \pi_{Zi})^{1-z}]^2Y_{rzi}(u)}{\Tilde{Y}_{rz}^2(u)}d\Lambda_{rz}(u)\\
&=& n_r\int_0^t\frac{d\Lambda_{rz}(u)}{\Tilde{W}_{rz}(u)} = \sigma_{rz}^2(t),\\
<U_{rz,\epsilon},U_{rz,\epsilon}>(t) &=& n_r\sum_{i=1}^{N_r}\int_0^tH_{rzi}^2(u)I[|H_{rzi}(u)| \geq \epsilon]Y_{rzi}(u)d\Lambda_{rz}(u) \overset{p}{\to} 0.
\end{eqnarray*}
Therefore by the martingale central limit theorem (see Theorem 5.1.1 in Fleming and Harrington, 1991), $\sqrt{n_r}\left\{ \Tilde{S}_{rz}(t) - S_{rz}(t) \right\} \overset{d}{\to} N \left( 0, S_{rz}^2(t)\sigma_{rz}^2(t) \right)$. Let $\Tilde{\Lambda}_{rz}(t) = \int_0^t\frac{d\Tilde{N}_{rz}(u)}{\Tilde{Y}_{rz}(u)}$, as $n_r \to \infty$,
\vspace*{-2ex}
\begin{eqnarray*}
\hat{\sigma}^2_{rz}(t) = \int_0^t\frac{n_rd\Tilde{N}_{rz}(u)}{\Tilde{W}_{rz}(u)(\Tilde{Y}_{rz}(u) - \Delta \Tilde{N}_{rz}(u))} \overset{p}{\to} n_r\int_0^t\frac{d\Tilde{\Lambda}_{rz}(u)}{\Tilde{W}_{rz}(u)}.
\end{eqnarray*}
By the corollary showed above, we have $\sup\limits_{0 \leq t \leq t^*}|\hat{\sigma}^2_{rz}(t) - \sigma_{rz}^2(t)| \overset{p}{\to} 0$.

\subsection*{B.4 Proof of Theorem 1}
\noindent By the consistency of $\Tilde{S}_{rz}(t)$ from Theorem 1*, we can obtain $\sup\limits_{0 \leq t \leq t^*}|\Tilde{S}_{rz}(t) - S_{rz}(t)| \overset{p}{\to} 0$. Then for any fixed $0 < t^* < \infty$,
\begin{eqnarray*}
|\Tilde{\mu}_{rz}(t^*) - \mu_{rz}(t^*)| &=& \left| \int_0^{t^*}\Tilde{S}_{rz}(t)dt - \int_0^{t^*}S_{rz}(t)dt \right|\\
&\leq& t^*\sup\limits_{0 \leq t \leq t^*}|\Tilde{S}_{rz}(t) - S_{rz}(t)| \\
&\overset{p}{\to}& 0.
\end{eqnarray*}
According to the limiting distribution of $\Tilde{S}_{rz}(t)$ derived in Theorem 1*, and using the asymptotic theory for functional of $\Tilde{S}_{rz}(t)$ (see Theorem 2.1 and Theorem 3.1 in Gill, 1983), we have
\vspace*{-2ex}
\begin{eqnarray*}
\sqrt{n_r}\left\{ \Tilde{\mu}_{rz}(t^*) - \mu_{rz}(t^*) \right\} \overset{d}{\to} N(0,\tau_{rz}^2(t^*)),
\end{eqnarray*}
where $\tau_{rz}^2(t^*) = n_r\int_0^{t^*}\{\int_u^{t^*}S_{rz}(t)dt\}^2\frac{d\Lambda_{rz}(u)}{\Tilde{W}_{rz}(u)}$, and by the consistency of $\Tilde{\Lambda}_{rz}(t)$, we have
\vspace{-2ex}
\begin{eqnarray*}
|\hat{\tau}_{rz}^2(t^*) - \tau_{rz}^2(t^*)| &\overset{p}{\to}& 0,
\end{eqnarray*}
where $\hat{\tau}_{rz}^2(t^*) = n_r\int_0^{t^*} \{ \int_u^{t^{*}}\Tilde{S}_{rz}(t)dt \}^2 \frac{d\Tilde{N}_{rz}(u)}{\Tilde{W}_{rz}(u)(\Tilde{Y}_{rz}(u) - \Delta \Tilde{N}_{rz}(u))}$.

\subsection*{B.5 Theorem 1 for IPSW}
\noindent The large sample properties in Theorem 1 are also valid for the IPSW-adjusted estimators. Most of the proofs are similar as shown for the CW-adjusted estimator, here we provide the distinctions when showing the consistency of $\Tilde{S}_{rz}(t)$. Lemma 1 is not required for the proof, but we make another assumption as follows: 
\vspace*{10pt}

\noindent \textbf{Assumption 5.} \textit{Let $d'_{ri}$ denote the inclusion in the target population, we assume that the marginal probability of being included in the target population is a constant, i.e., $E[P(d'_{ri}|X_{ri})] = C_r$. In addition, the conditional probability of being included in the target population, $P(d'_{ri}|X_{ri})$, is independent with the observed time $Y_{ri}$.}
\vspace*{10pt}

\noindent In Web Appendix B.3, we show that for any fixed $0 < t^* < \infty$,
\begin{eqnarray*}
P \left( \frac{S_{rz}(t) - \Tilde{S}_{rz}(t)}{S_{rz}(t)} = V_{rz}(t), t \in [0,t^*] \right) \to 1,
\end{eqnarray*}
where
\vspace*{-2ex}
\begin{eqnarray*}
V_{rz}(t) &=& \int_0^t\frac{\Tilde{S}_{rz}(u-)}{S_{rz}(u)}\frac{I[\Tilde{Y}_{rz}(u) > 0]}{\Tilde{Y}_{rz}(u)}d\Tilde{M}_{rz}(u) \\
&=& \sum_{i=1}^{N_r}\int_0^t\frac{\Tilde{S}_{rz}(u-)}{S_{rz}(u)}\frac{I[\Tilde{Y}_{rz}(u) > 0]}{\Tilde{Y}_{rz}(u)}d_{ri}\gamma_{ri}\hat{q}_{rzi}dM_{rzi}(u) \\
&=& \frac{1}{N_r}\sum_{i=1}^{N_r}\int_0^t\frac{\Tilde{S}_{rz}(u-)}{S_{rz}(u)}\frac{I[\Tilde{Y}_{rz}(u) > 0]}{\Tilde{Y}_{rz}(u)/N_r}\frac{d_{ri}}{\rho_r(\bX_{ri})}P(d'_{ri}=1|\bX_{ri})\hat{q}_{rzi}dM_{rzi}(u) \\
&\overset{p}{\to}& \frac{1}{N_r}\sum_{i=1}^{N_r}\int_0^t\frac{\Tilde{S}_{rz}(u-)}{S_{rz}(u)}\frac{I[\bar{Y}_{rz}(u) > 0]}{\omega_{rz}(u)C_r}\hat{q}_{rzi}\frac{d_{ri}}{\rho_r(\bX_{ri})}P(d'_{ri}=1|\bX_{ri})dM_{rzi}(u) \\
&\overset{p}{\to}& E\left[ \int_0^t\frac{\Tilde{S}_{rz}(u-)}{S_{rz}(u)}\frac{I[\bar{Y}_{rz}(u) > 0]}{\omega_{rz}(u)}dM_{rzi}(u)\right],
\end{eqnarray*}
where $\bar{Y}_{rz}(u) = \sum_{i=1}^{n_r}\hat{q}_{rzi}Y_{rzi}(u)$. And by Assumption 5, we have $\Tilde{Y}_{rz}(u)/N_r = \frac{1}{N_r}\sum_{i=1}^{N_r}d_{ri}\gamma_{ri}\hat{q}_{rzi}\\I[Y_{rzi} \geq u] = \frac{1}{N_r}\sum_{i=1}^{N_r}\frac{d_{ri}}{\rho_r(X_i)}P(d'_{ri}=1|\bX_{ri})\hat{q}_{rzi}I[Y_{rzi} \geq u] \overset{p}{\to} P(Y_{rzi} \geq u) C_r = \omega_{rz}(u) C_r$. The remaining of the proofs are similar as for CW-adjusted estimator in Theorem 1.

\section*{Web Appendix C: Proofs of Theorem 2}
\noindent We show the consistency of the weighted G-formula estimators by using CW $\hat{p}_{ri}$ and IPSW $\hat{\gamma}_{ri}$. The asymptotic distribution of the CW-adjusted and IPSW-adjusted G-formula estimators can be derived by the M-estimator theory (Stefanski and Boos, 2002) and the Delta method (Dowd, Greene, and Norton, 2014). Let the outcome models for $z_r \in \{0,1\}$ in Region $r$ be $m_{r0}(\bX_r) = \phi^{-1}(\hat{\beta}_{r0} + \hat{\bbeta}_{r2}\bg(\bX_r^T))$ and $m_{r1}(\bX_r) = \phi^{-1}(\hat{\beta}_{r0} + \hat{\beta}_{r1} + \hat{\bbeta}_{r2}\bg(\bX_r^T) + \hat{\bbeta}_{r3}\bg(\bX_r^T))$. Here, we assume that the outcome models are not misspecified. We re-write $\hat{\Delta}_r^{GF}(t^*)$ as follows:
\vspace*{-2ex}
\begin{eqnarray*}
    \hat{\Delta}_r^{GF}(t^*) = \hat{\mu}_{r1}^{GF}(t^*) - \hat{\mu}_{r0}^{GF}(t^*),
\end{eqnarray*}
where $\hat{\mu}_{r1}^{GF}(t^*) = \frac{\sum_{i=1}^{n_r} \hat{\xi}_{ri}m_{r1}(\bX_{ri})}{\sum_{i=1}^{n_r} \hat{\xi}_{ri}}$ and $\hat{\mu}_{r0}^{GF}(t^*) = \frac{\sum_{i=1}^{n_r} \hat{\xi}_{ri}m_{r0}(\bX_{ri})}{\sum_{i=1}^{n_r} \hat{\xi}_{ri}}$.

\subsection*{C.1 Consistency of CW-adjusted G-formula estimator}
For the calibration weights $\hat{p}_{ri}$,
\vspace*{-2ex}
\begin{eqnarray*}
    \hat{\mu}_{r1}^{GF}(t^*) &=& \frac{\sum_{i=1}^{n_r} \hat{p}_{ri}m_{r1}(\bX_{ri})}{\sum_{i=1}^{n_r} \hat{p}_{ri}} \\
    &=& \phi^{-1}(\hat{\beta}_{r0} + \hat{\beta}_{r1} + \hat{\bbeta}_{r2}\Tilde{\bg}^T + \hat{\bbeta}_{r3}\Tilde{\bg}^T),
\end{eqnarray*}
where $\Tilde{\bg}$ is the sample moment estimates of $\bg(\bX)$ from the target population. The equation on the second line is obtained by the constraint function in Equation 3.4. Therefore,
\vspace*{-2ex}
\begin{eqnarray*}
    E_{F^*}[\hat{\mu}_{r1}^{GF}(t^*)]  &\overset{p}{\to}& E_{F^*}[\phi^{-1}(\beta_{r0} + \beta_{r1} + \bbeta_{r2}\bg(\bX^T) + \bbeta_{r3}\bg(\bX^T))] \\
    &=& E_{F^*}[m_{r1}(\bX)] \\
    &=& \mu_{r1}(t^*).
\end{eqnarray*}
Here, the expectation is taken on the distribution in the target population. Similarly, we have
\vspace*{-2ex}
\begin{eqnarray*}
    E_{F^*}[\hat{\mu}_{r1}^{GF}(t^*)]  &\overset{p}{\to}& E_{F^*}[\phi^{-1}(\beta_{r0} + \bbeta_{r2}\bg(\bX^T))] \\
    &=& E_{F^*}[m_{r0}(\bX)] \\
    &=& \mu_{r0}(t^*).
\end{eqnarray*}
Then, 
\vspace*{-2ex}
\begin{eqnarray*}
    E_{F^*}[\hat{\Delta}_r^{GF}(t^*)]  &\overset{p}{\to}&  \mu_{r1}(t^*) - \mu_{r0}(t^*) = \Delta_r(t^*).
\end{eqnarray*}

\subsection*{C.2 Consistency of IPSW-adjusted G-formula estimator}
\noindent We use the M-estimator theory to show the consistency of the IPSW-adjusted G-formula estimator. Let $\btheta_r^{GF} = [\theta_{r0}^{GF}, \theta_{r1}^{GF}]^T$ as the collection of parameters to be estimated. Then $\hat{\Delta}_r^{GF}(t^*) = \hat{\theta}_{r1}^{GF} - \hat{\theta}_{r0}^{GF}$ jointly solves the estimation equations as follows:
\vspace*{-2ex}
\begin{eqnarray*}
    \sum_{i=1}^{n_r}\bPhi_{ri}^{GF}(\btheta_r^{GF}) = \sum_{i=1}^{n_r} \begin{pmatrix}
\hat{\gamma}_{ri}\{m_{r1}(\bX_{ri}) - \theta_{r1}^{GF}\} \\
\hat{\gamma}_{ri}\{m_{r0}(\bX_{ri}) - \theta_{r0}^{GF}\} 
\end{pmatrix} = 0.
\end{eqnarray*}
Taking expectation of $\bPhi_{ri}^{GF}(\btheta_r^{GF})$ on the distribution of $F_r$, we have
\vspace*{-2ex}
\begin{eqnarray*}
    E_{F_r}\bPhi_{ri}^{GF}(\btheta_r^{GF}) &=& \begin{pmatrix}
E_{F_r}[\hat{\gamma}_{ri}\{m_{r1}(\bX_{ri}) - \theta_{r1}^{GF}\}] \\
E_{F_r}[\hat{\gamma}_{ri}\{m_{r0}(\bX_{ri}) - \theta_{r0}^{GF}\}] 
\end{pmatrix} \\
&=& \begin{pmatrix}
E_{F_r}[\frac{dF^*(\bX_{ri})}{dF_r(\bX_{ri})}\{m_{r1}(\bX_{ri}) - \theta_{r1}^{GF}\}] \\
E_{F_r}[\frac{dF^*(\bX_{ri})}{dF_r(\bX_{ri})}\{m_{r0}(\bX_{ri}) - \theta_{r0}^{GF}\}] 
\end{pmatrix} \\
&=& \begin{pmatrix}
E_{F^*}[m_{r1}(\bX_{ri}) - \theta_{r1}^{GF}] \\
E_{F^*}[m_{r0}(\bX_{ri}) - \theta_{r0}^{GF}] 
\end{pmatrix} \\
&=& \begin{pmatrix}
E_{F^*}[m_{r1}(\bX_{ri})] - \theta_{r1}^{GF} \\
E_{F^*}[m_{r0}(\bX_{ri})] - \theta_{r0}^{GF} 
\end{pmatrix}.
\end{eqnarray*}
By solving $ E_{F_r}\bPhi_{ri}^{GF}(\btheta_r^{GF}) = 0$, we have $\theta_{r1}^{GF} = E_{F^*}[m_{r1}(\bX_{ri})]$ and $\theta_{r0}^{GF} = E_{F^*}[m_{r0}(\bX_{ri})]$. As such, $\hat{\Delta}_r^{GF}(t^*)$ is consistent with $\theta_{r1}^{GF} - \theta_{r0}^{GF} = E_{F^*}[m_{r1}(\bX_{ri})] - E_{F^*}[m_{r0}(\bX_{ri})] = \Delta_r(t^*)$.

\section*{Web Appendix D: Proofs of Theorems 3 and 4}
\noindent The large sample properties for the weighted Hajek and weighted Augmented estimators are proved by the M-estimator theory.

\subsection*{D.1 Theorem 3}
\noindent We re-write the weighted Hajek estimator, $\hat{\Delta}_r^{HJ}(t^*)$, as follows:
\vspace*{-2ex}
\begin{eqnarray*}
    \hat{\Delta}_r^{HJ}(t^*) = \hat{\mu}_{r1}^{HJ}(t^*) - \hat{\mu}_{r0}^{HJ}(t^*),
\end{eqnarray*}
where $\hat{\mu}_{r1}^{HJ}(t^*) = \frac{\sum_{i=1}^{n_r}\hat{\xi}_{ri}\hat{q}_{r1i}w_{ri}Y_{ri}}{\sum_{i=1}^{n_r}\hat{\xi}_{ri}\hat{q}_{r1i}w_{ri}}$ and $\hat{\mu}_{r0}^{HJ}(t^*) = \frac{\sum_{i=1}^{n_r}\hat{\xi}_{ri}\hat{q}_{r0i}w_{ri}Y_{ri}}{\sum_{i=1}^{n_r}\hat{\xi}_{ri}\hat{q}_{r0i}w_{ri}}$. Let $\btheta_r^{HJ} =$ \\ $[\theta_{r0}^{HJ}, \theta_{r1}^{HJ}]^T$ as the collection of parameters to be estimated. Then $\hat{\Delta}_r^{HJ}(t^*) = \\ \hat{\theta}_{r1}^{HJ} - \hat{\theta}_{r0}^{HJ}$ jointly solves the estimation equations as follows:
\vspace*{-2ex}
\begin{eqnarray*}
    \sum_{i=1}^{n_r}\bPhi_{ri}^{HJ}(\btheta_r^{HJ}) = \sum_{i=1}^{n_r} \begin{pmatrix}
\hat{\xi}_{ri}\hat{q}_{r1i}w_{ri}\{Y_{ri} - \theta_{r1}^{HJ}\} \\
\hat{\xi}_{ri}\hat{q}_{r0i}w_{ri}\{Y_{ri} - \theta_{r0}^{HJ}\} 
\end{pmatrix} = 0.
\end{eqnarray*}

\noindent First, for the calibration weight (i.e., $\hat{\xi}_{ri} = \hat{p}_{ri}$), by Lemma 1, we have:
\vspace*{-2ex}
\begin{eqnarray*}
    \sum_{i=1}^{n_r}\hat{p}_{ri}\hat{q}_{r1i}w_{ri}\{Y_{ri} - \theta_{r1}^{HJ}\}  &\overset{p}{\to}& \frac{1}{N_r}\sum_{i=1}^{N_r}\frac{d_{ri}}{\rho_r(\bX_{ri})}w_{ri}\{Y_{ri} - \theta_{r1}^{HJ}\} \\
     &\overset{p}{\to}& E[\frac{d_{ri}}{\rho_r(\bX_{ri})}\frac{\delta_i^*}{\hat{G}(Y_i)}\{Y_{ri} - \theta_{r1}^{HJ}\}] \\
     &\overset{p}{\to}& E[Y_{ri}] -\theta_{r1}^{HJ}.
\end{eqnarray*}
By solving $E[Y_{ri}] -\theta_{r1}^{HJ} = 0$, we have $\hat{\theta}_{r1}^{HJ} = E[Y_{ri}] = \mu_{r1}(t^*)$. So that $\hat{\theta}_{r1}^{HJ}$ is consistent with $\mu_{r1}(t^*)$. Similarly, we can show $\hat{\theta}_{r0}^{HJ}$ is consistent with $\mu_{r0}(t^*)$. Therefore, $\hat{\Delta}_r^{HJ}(t^*)$ is consistent with $\Delta_r(t^*)$ under the calibration weight.
\vspace*{10pt}

\noindent Second, for the inverse probability of sampling weight (i.e., $\hat{\xi_{ri}} = \hat{\gamma}_{ri}$), under the Assumption 5, we have:
\vspace*{-2ex}
\begin{eqnarray*}
    \sum_{i=1}^{n_r}\hat{\gamma}_{ri}\hat{q}_{r1i}w_{ri}\{Y_{ri} - \theta_{r1}^{HJ}\}  &\overset{p}{\to}& \frac{1}{N_r}\sum_{i=1}^{N_r}\frac{d_{ri}}{\rho_r(\bX_{ri})}P(d'_{ri}=1|\bX_{ri})w_{ri}\{Y_{ri} - \theta_{r1}^{HJ}\} \\
     &\overset{p}{\to}& E[\frac{d_{ri}}{\rho_r(\bX_{ri})}\frac{\delta_i^*}{\hat{G}(Y_i)}C_r\{Y_{ri} - \theta_{r1}^{HJ}\}] \\
     &\overset{p}{\to}& E[Y_{ri}] -\theta_{r1}^{HJ}.
\end{eqnarray*}
Therefore, $\hat{\Delta}_r^{HJ}(t^*)$ is consistent with $\Delta_r(t^*)$ under the inverse probability of sampling weight.
\vspace*{10pt}

\noindent By the M-estimator theory, as $n_r \to \infty$, $\sqrt{n_r}(\hat{\btheta}_r^{HJ}-\btheta_r^{HJ}) \overset{d}{\to} N(0, n_r\Sigma_r^{HJ})$, where $\Sigma_r^{HJ}$ is the sandwich variance estimator for $\hat{\btheta}_r^{HJ}$ as:
\begin{eqnarray*}
    \Sigma_r^{HJ} = \left\{ \sum_{i=1}^{n_r} \frac{\partial \bPhi_{ri}^{HJ}(\btheta_r^{HJ})}{(\partial \hat{\btheta}_r^{HJ})^T} \right\}^{-1} \left\{ \sum_{i=1}^{n_r} \bPhi_{ri}^{HJ}(\hat{\btheta}_r^{HJ}) \bPhi_{ri}^{HJ}(\hat{\btheta}_r^{HJ})^T \right\} \left\{  \sum_{i=1}^{n_r} \frac{\partial \bPhi_{ri}^{HJ}(\btheta_r^{HJ})}{(\partial \hat{\btheta}_r^{HJ})^T} \right\}^{-1}.
\end{eqnarray*}
Then by continuous mapping theorem, $\sqrt{n_r}\left\{ \hat{\Delta}_r^{HJ}(t^*) - \Delta_r(t^*) \right\} \overset{d}{\to} N \left ( 0,\sigma_{r,HJ}^2(t^*) \right )$. The asymptotic variance $\sigma_{r,HJ}^2(t^*)$ can be estimated by $n_r[1,-1]\Sigma_r^{HJ}[1,-1]^T$.

\subsection*{D.2 Theorem 4}
\noindent The proofs of Theorem 4 are similar in Theorem 3 based on the M-estimator theory, and the consistency of $\hat{\Delta}_r^{AG}(t^*)$ is ensured by the consistency of $\hat{\Delta}_r^{GF}(t^*)$ and $\hat{\Delta}_r^{HJ}(t^*)$. Here, we show that the weighted Augmented estimator is also consistent with $\Delta_r(t^*)$ when the outcome models are mis-specified.
\vspace*{10pt}

\noindent We re-write the weighted Augmented estimator, $\hat{\Delta}_r^{AG}(t^*)$, as follows:
\vspace*{-2ex}
\begin{eqnarray*}
    \hat{\Delta}_r^{AG}(t^*) = \hat{\mu}_{r1}^{AG}(t^*) - \hat{\mu}_{r0}^{AG}(t^*),
\end{eqnarray*}
where $\hat{\mu}_{r1}^{AG}(t^*) = \frac{\sum_{i=1}^{n_r}\hat{\xi}_{ri}\hat{q}_{r1i}w_{ri}\{Y_{ri}-m_{r1}(\bX_{ri})\}}{\sum_{i=1}^{n_r}\hat{\xi}_{ri}\hat{q}_{r1i}w_{ri}} + \frac{\sum_{i=1}^{n_r} \hat{\xi}_{ri}m_{r1}(\bX_{ri})}{\sum_{i=1}^{n_r} \hat{\xi}_{ri}}$ and \\
$\hat{\mu}_{r0}^{AG}(t^*) = \frac{\sum_{i=1}^{n_r}\hat{\xi}_{ri}\hat{q}_{r0i}w_{ri}\{Y_{ri}-m_{r0}(\bX_{ri})\}}{\sum_{i=1}^{n_r}\hat{\xi}_{ri}\hat{q}_{r0i}w_{ri}} + \frac{\sum_{i=1}^{n_r} \hat{\xi}_{ri}\hat{q}_{r0i}m_{r0}(\bX_{ri})}{\sum_{i=1}^{n_r} \hat{\xi}_{ri}\hat{q}_{r0i}}$.
\vspace*{10pt}

\noindent First, for the calibration weight (i.e., $\hat{\xi_{ri}} = \hat{p}_{ri}$), we have
\vspace*{-2ex}
\begin{eqnarray*}
    \hat{\mu}_{r1}^{AG}(t^*) &\overset{p}{\to}& \frac{\sum_{i=1}^{n_r}\hat{p}_{ri}\hat{q}_{r1i}w_{ri}Y_{ri}}{\sum_{i=1}^{n_r}\hat{p}_{ri}\hat{q}_{r1i}w_{ri}} + 
    \sum_{i=1}^{n_r}\hat{p}_{ri}m_{r1}(\bX_{ri}) - \frac{\sum_{i=1}^{n_r}\hat{p}_{ri}w_{ri}m_{r1}(\bX_{ri})}{\sum_{i=1}^{n_r}\hat{p}_{ri}w_{ri}} \\
    &=& \hat{\mu}_{r1}^{HJ}(t^*) + \sum_{i=1}^{n_r} \left\{ \hat{p}_{ri} - \frac{\hat{p}_{ri}w_{ri}}{\sum_{i=1}^{n_r}\hat{p}_{ri}w_{ri}}  \right\}m_{r1}(\bX_{ri}) \\
    &=& \hat{\mu}_{r1}^{HJ}(t^*) + \sum_{i=1}^{n_r} \hat{p}_{ri} \left\{ 1 - \frac{w_{ri}}{\sum_{i=1}^{n_r}\hat{p}_{ri}w_{ri}}  \right\}m_{r1}(\bX_{ri}). 
\end{eqnarray*}
Since $w_{ri}/\sum_{i=1}^{n_r}\hat{p}_{ri}w_{ri} \overset{p}{\to} 1$, then we have $\hat{\mu}_{r1}^{AG}(t^*) \overset{p}{\to}  \hat{\mu}_{r1}^{HJ}(t^*)$. Similarly, we can show that $\hat{\mu}_{r0}^{AG}(t^*) \overset{p}{\to}  \hat{\mu}_{r0}^{HJ}(t^*)$. Therefore, we see that $\hat{\Delta}_r^{AG}(t^*)$ is consistent with $\Delta_r(t^*)$ regardless of the outcome model specification under the CW.
\vspace*{10pt}

\noindent Second, for the inverse probability of sampling weight (i.e., $\hat{\xi_{ri}} = \hat{\gamma}_{ri}$), we have
\begin{eqnarray*}
    \hat{\mu}_{r1}^{AG}(t^*) &\overset{p}{\to}& \hat{\mu}_{r1}^{HJ}(t^*) - \frac{\sum_{i=1}^{n_r}\hat{\gamma}_{ri}\hat{q}_{r1i}w_{ri}m_{r1}(\bX_{ri})}{\sum_{i=1}^{n_r}\hat{\gamma}_{ri}\hat{q}_{r1i}w_{ri}} + \frac{\sum_{i=1}^{n_r}\hat{\gamma}_{ri}m_{r1}(\bX_{ri})}{\sum_{i=1}^{n_r}\hat{\gamma}_{ri}}. 
\end{eqnarray*}
By M-estimator theory (see proof in Web Appendix C.2), we can show that
\begin{eqnarray*}
\frac{\sum_{i=1}^{n_r}\hat{\gamma}_{ri}\hat{q}_{r1i}w_{ri}m_{r1}(\bX_{ri})}{\sum_{i=1}^{n_r}\hat{\gamma}_{ri}\hat{q}_{r1i}w_{ri}} &\overset{p}{\to}& E_{F^*}(m_{r1}(\bX_{ri})), \\
    \frac{\sum_{i=1}^{n_r}\hat{\gamma}_{ri}m_{r1}(\bX_{ri})}{\sum_{i=1}^{n_r}\hat{\gamma}_{ri}} &\overset{p}{\to}& E_{F^*}(m_{r1}(\bX_{ri})).
\end{eqnarray*}
As such, $\hat{\mu}_{r1}^{AG}(t^*) \overset{p}{\to}  \hat{\mu}_{r1}^{HJ}(t^*)$, and similarly, $\hat{\mu}_{r0}^{AG}(t^*) \overset{p}{\to}  \hat{\mu}_{r0}^{HJ}(t^*)$. Therefore, $\hat{\Delta}_r^{AG}(t^*)$ is consistent with $\Delta_r(t^*)$ regardless of the outcome model specification under the IPSW.

\section*{Web Appendix E: Additional Simulation Set-Ups and Results}

\subsection*{E.1 True Average Region-Specific RMST and RMST Difference}
\noindent The average region-specific RMST at $t^*$, $\mu_{rz}(t^*)$, is averaged over the distribution of $\bX$ from the target population represented by a common distribution $F^*:$ $X_1 \sim Unif(0,1)$ and $X_2 \sim N(1,1)$. Its true value $\bar{\mu}_{rz}(t^*)$ is
\vspace{-2ex}
\begin{eqnarray*}
    \bar{\mu}_{rz}(t^*) &=& \int \mu_{rz}(t^*|\bx) dF^*(\bx) \\
    &=& \int \int_0^{t^*} S_{rz}(t|\bx) dF^*(\bx) \\
    &=& \int \int_0^{t^*} exp\{ -\int_0^t h(u|z,r,\bx) \} dF^*(\bx).
\end{eqnarray*}
where $h(u|z,r,\bx) = h(u|Z=z,R=r,X_1=x_1,X_2=x_2)$ is the hazard function of Equation 5.3. The true region-specific average RMST difference is $\bar{\Delta}_r(t^*) = \bar{\mu}_{r1}(t^*) - \bar{\mu}_{r0}(t^*)$.

\subsection*{E.2 Parameters in the Sampling Score Models}
\begin{table}[H]
\caption{Values of parameters in the sampling score models, Equations 5.1 and 5.2 under four sampling scenarios. Scenario 1: Log-linear sampling (Equation 5.1) with moderate SMDs; Scenario 2: Log-linear sampling with large SMDs; Scenario 3: Logistic-nonlinear sampling (Equation 5.2) with moderate SMDs; Scenario 4: Logistic-nonlinear sampling with large SMDs.}
\label{tabC.1}
\begin{center}
\begin{tabular}{ccccccc}
\hline
Equation 5.1 &\multicolumn{3}{c}{Scenario 1} & \multicolumn{3}{c}{Scenario 2} \\
\hline
Region&$\eta_{r0}$&$\eta_{r1}$&$\eta_{r2}$&$\eta_{r0}$&$\eta_{r1}$&$\eta_{r2}$\\
\hline
$r=1$ & -5.0 & 0.8 & 0.30 & -5.0 & 2.5 & 0.50 \\
$r=2$ & -5.0 & 0.7 & 0.27 & -5.0 & 2.3 & 0.55 \\
$r=3$ & -5.0 & 0.6 & 0.25 & -5.0 & 2.0 & 0.60 \\
\hline
Equation 5.2 &\multicolumn{3}{c}{Scenario 3} & \multicolumn{3}{c}{Scenario 4} \\
\hline
Region&$\eta_{r0}^*$&$\eta_{r1}^*$&$\eta_{r2}^*$&$\eta_{r0}^*$&$\eta_{r1}^*$&$\eta_{r2}^*$\\
\hline
$r=1$ & -3.0 & 0.6 & -0.15 & -2.3 & 3.0 & -0.20 \\
$r=2$ & -3.0 & 0.5 & -0.10 & -2.3 & 2.5 & -0.15 \\
$r=3$ & -3.0 & 0.4 & -0.05 & -2.3 & 2.0 & -0.10 \\
\hline
\end{tabular}
\end{center}
\end{table}

\subsection*{E.3 Absolute Standardized Mean Differences of $X_1$ and $X_2$}

\begin{table}[H]
\caption{Absolute standardized mean differences of $X_1$ and $X_2$ between the enrolled patients in each region and the target population in simulation study. $X_1 \sim Unif(0,1)$ and $X_2 \sim N(1,1)$ in the target population. We generate a large data with sample size of 90,000, 30,000 individuals in each region, under four scenarios of the sampling models.}
\label{tabC.2}
\begin{center}
\begin{tabular}{cccccc}
\hline
&Region& Scenario 1 & Scenario 2 & Scenario 3 & Scenario 4 \\
\hline
$X_1$ & $r=1$ & 0.235 & 0.692 & 0.208 & 0.517 \\
& $r=2$ & 0.207 & 0.656 & 0.164 & 0.497 \\
& $r=3$ & 0.161 & 0.564 & 0.126 & 0.481 \\
\hline
$X_2$ & $r=1$ & 0.305 & 0.502 & 0.297 & 0.696 \\
& $r=2$ & 0.276 & 0.545 & 0.245 & 0.683 \\
& $r=3$ & 0.238 & 0.609 & 0.199 & 0.647 \\
\hline
\end{tabular}
\end{center}
\end{table}

\subsection*{E.4 Simulation Results for Regions 2 and 3}
\begin{figure}[H]
\centering\includegraphics[scale = 0.5]{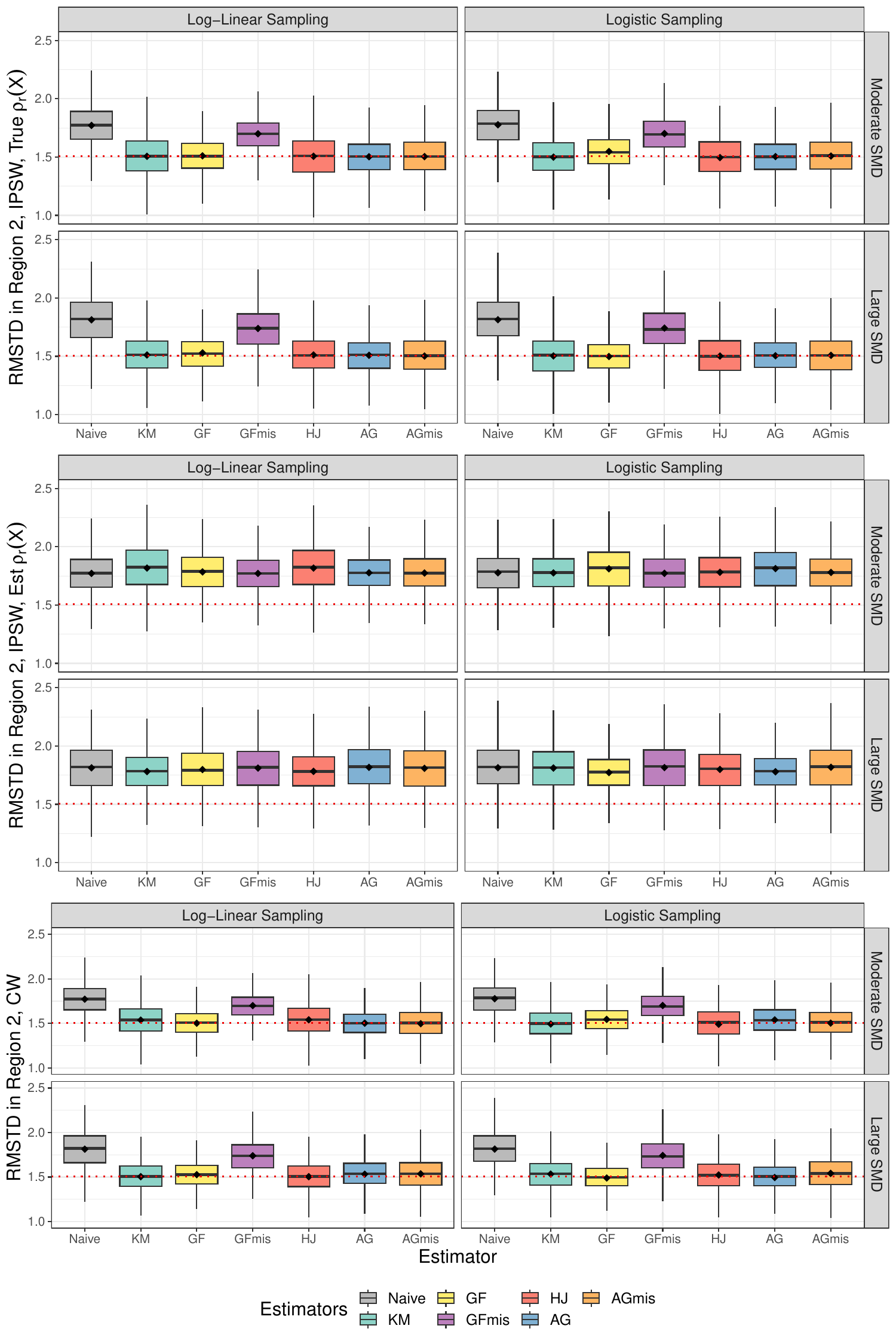}
\caption[Boxplots of average RMSTD in Region 2 in simulation study.]{Boxplots of estimated average RMST difference (RMSTD) in Region 2 under four sampling scenarios in simulation study. Upper panel: IPSW-adjusted estimators with true sampling score; Middle panel: IPSW-adjusted estimators with estimated sampling score; Bottom panel: CW-adjusted estimators.}
\end{figure}

\begin{figure}[H]
\centering\includegraphics[scale = 0.5]{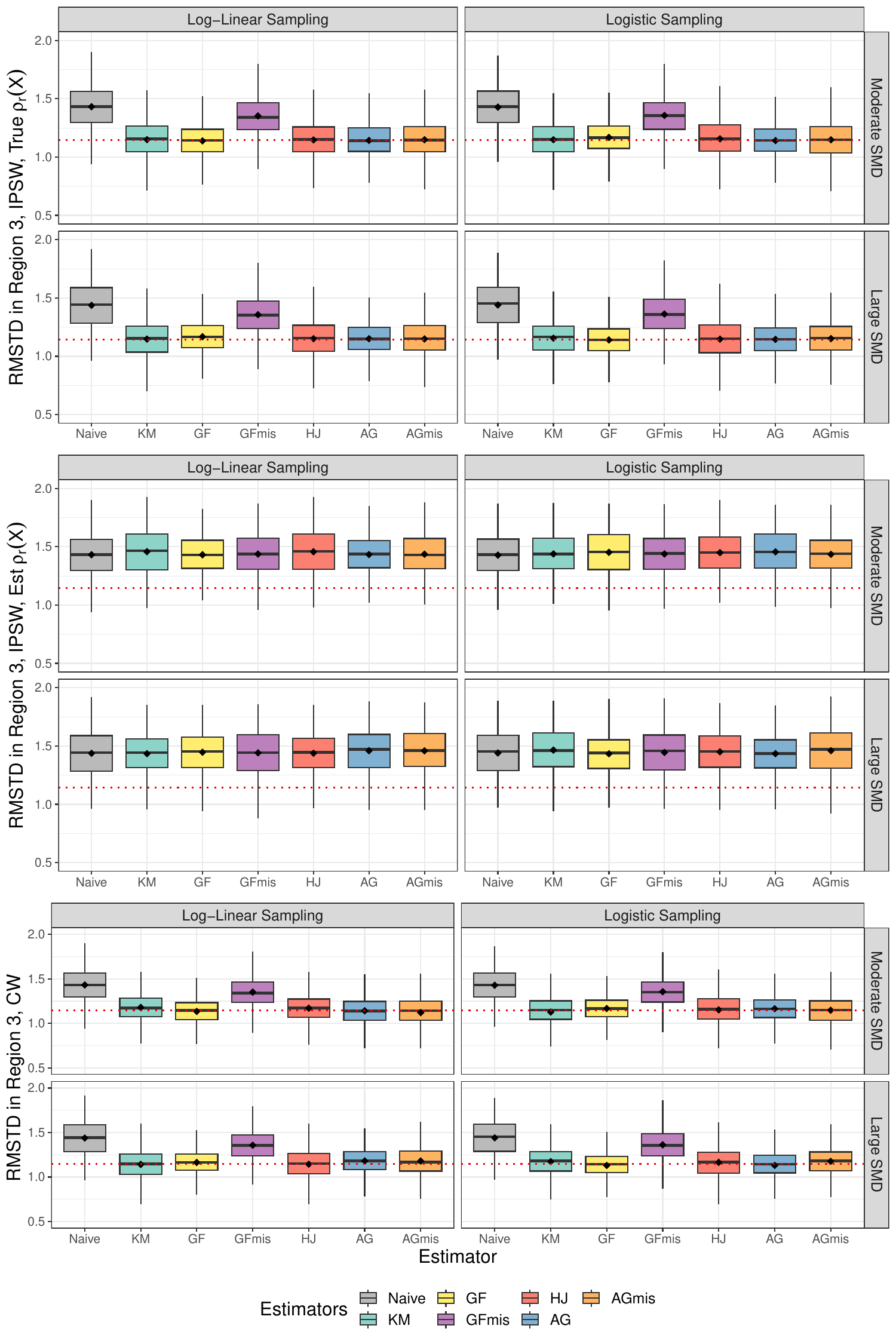}
\caption[Boxplots of average RMSTD in Region 3 in simulation study.]{Boxplots of estimated average RMST difference (RMSTD) in Region 3 under four sampling scenarios in simulation study. Upper panel: IPSW-adjusted estimators with true sampling score; Middle panel: IPSW-adjusted estimators with estimated sampling score; Bottom panel: CW-adjusted estimators.}
\end{figure}

\section*{Web F: Additional Case Study Results for PLATO Trial}
\subsection*{F.1: Variable Selection}
\noindent Previous effect modifier analyses suggested that the maintenance aspirin dosage is the only variable that explained the statistically significant regional interaction effect with a p-value below 0.05 (Mahaffey et al., 2011; K.J. Carroll and Fleming, 2013). We conduct a univariable effect modifier analysis with the characteristics listed in Mahaffey et al. (2011) to select more covariates for the weighting methods. We fit two univariable models in the pooled dataset for each candidate variable: 1) IPCW RMST regression model (Tian, Zhao, and Wei, 2014) with truncation time $t^*=360$ days; 2) Cox regression model (Cox, 1972). Each model includes a treatment indicator, a single covariate, and a treatment-by-covariate interaction. A covariate will be selected if the p-value of its interaction with treatment is less than 0.2 in either of the two models. We choose the significance level of 0.2 to enhance the power to detect significant effect modifiers.
\vspace*{10pt}

\noindent As a result, seven variables are selected as potential effect modifiers, including weight ($\geq 80$ kg), hypertension, percutaneous coronary intervention, smoking (ever), peripheral arterial disease, Killip classification (Level I vs. Level II-IV), and maintenance aspirin dosage. The first six variables are binary, and we categorize maintenance aspirin dosage into three levels (Mahaffey et al., 2011): 1) $\leq 100$ mg, 2) $(100,300)$ mg, and 3) $\geq 300$ mg, and we let the first level be the reference group. Overall, we have eight binary covariates included in the weighting methods.

\subsection*{F.2: Results from Four-Region Analysis}
\noindent Web Figure 3 and Web Table 3 present the estimated average RMST differences and the associated 95\% CIs comparing ticagrelor and clopidogrel for the primary outcome among four regions. The unadjusted RMST differences are 6.6 days (95\% CI: -3.1, 16.3) in Asia and Australia, 6.2 days (95\% CI: -6.7, 19) in Central and South America, 5.5 days (95\% CI: 2.6, 8.4) in Europe, Middle East, and Africa, and -4.1 days (95\% CI: -12.5, 4.4) in North America. The unadjusted results indicate that ticagrelor is numerically, though not statistically significantly, more effective than clopidogrel in Asia and Australia, and Central and South America, and is significantly more effective in Europe, the Middle East, and Africa. However, it is numerically less effective than clopidogrel in North America. The consistency test for the Naive estimator shows that treatment effects are not heterogeneous across four regions (p = 0.20). 
\vspace*{10pt}

\noindent In Europe, Middle East, and Africa, the dominant region, the estimated RMST differences are very similar across four estimators within the same weighting method. The CW-adjusted estimators yield slightly higher RMST differences than the IPSW-adjusted estimators. Both weighting methods indicate that ticagrelor is significantly more effective than clopidogrel in this region. Under both weighting approaches, the weighted G-formula estimators have the largest variance. In North America, the average RMST differences from all weighted estimators are close to 0, indicating that ticagrelor has the same effect as clopidogrel in North America after balancing the distributions of the selected variables across four regions. However, the IPSW and CW methods yield opposite results in the other two regions. For example, the average RMST difference in Central and South America from all IPSW-adjusted estimators is positive, while it is negative from all CW-adjusted estimators. The results from Asia and Australia show a similar pattern. This discrepancy is potentially due to misspecification of the estimated sampling score in the IPSW-adjusted estimators. 
\vspace*{10pt}

\noindent The consistency tests for most CW-adjusted and IPSW-adjusted estimators reveal no regional treatment effect heterogeneity (see p-values in Web Table 3). However, the test using the CW-adjusted Augmented estimators demonstrates that the treatment effects are significantly heterogeneous across regions (p = 0.05). The global RMST differences from other estimators are summarized in Web Table 3.
\vspace*{10pt}

\begin{table}[ht]
\caption[Average RMSTD in four-region analysis for PLATO trial.]{Estimated average RMST differences with 95\% CIs at $t^*=360$ days from 1) A\&A: Asia and Australia, 2) C\&SA: Central and South America, 3) EU,ME\&A: Europe, Middle East and Africa, and 4) NA: North America in the four-region analysis, and the results from the regional consistency test with p-values and global RMST difference.}
\label{tabC.3}
\begin{center}
\small
\begin{tabular}{|c|cccccc|}
\hline
Method & \multicolumn{4}{c}{RMSTD (95\% CI) in} & P-Val & Global \\
& A\&A & C\&SA & EU,ME\&A & NA & & RMSTD \\
\hline
Naive & 6.6 & 6.2 & 5.5 & -4.1 & 0.20 & 4.7 \\
& (-3.1, 16.3)& (-6.7, 19)& (2.6, 8.4)& (-12.5, 4.4)&& (2.0, 7.3) \\
IPSW.KM & 6.6 & 6.1 & 5.2 & 0.3 & 0.83 & 5.0 \\
& (-6.9, 20.1) & (-6, 18.2) & (2.4, 7.9) & (-10.1, 10.7) && (2.4, 7.5) \\
IPSW.GF & -2.9 & 2.7 & 5.8 & 1.6 & 0.76 & 5.2 \\
& (-21.7, 16) & (-14.8, 20.2) & (2.2, 9.5) & (-12.5, 15.6) && (1.8, 8.6) \\
IPSW.HJ & 5.9 & 4.5 & 5.2 & -0.2 & 0.82 & 4.9 \\
& (-6.1, 17.8) & (-7.2, 16.1) & (2.4, 8.1) & (-11.3, 10.8) && (2.3, 7.5) \\
IPSW.AG & 2.0  & 2.6 & 5.8 & 0.7 & 0.74 & 5.2 \\
& (-11.1, 15.1) & (-9.1, 14.2) & (3.0, 8.7) & (-10.4, 11.8) && (2.6, 7.8) \\
CW.KM & -5.0 & -0.9 & 4.9 & 1.2 & 0.43 & 3.9 \\
& (-18.9, 9.0) & (-13.6, 11.9) & (1.8, 7.9) & (-8.7, 11.0) && (1.1, 6.7) \\
CW.GF & -11.0 & -3.0 & 6.6 & 1.6 & 0.24 & 5.1 \\
& (-30.3, 8.2) & (-21.5, 15.5) & (2.4, 10.8) & (-12.0, 15.3) && (1.3, 8.9) \\
CW.HJ & -6.1 & -1.3 & 5.3 & 0.8 & 0.33 & 4.0 \\
& (-20.3, 8.0) & (-14.6, 12.1) & (1.8, 8.7) & (-9.5, 11.1) && (0.9, 7.1) \\
CW.AG & -10.8 & -2.3 & 6.5 & 2.4 & 0.05 & 4.7 \\
& (-23.9, 2.3) & (-15.9, 11.2) & (3.1, 10.0) & (-7.9, 12.7) && (1.6, 7.8) \\
\hline
\end{tabular}
\end{center}
\end{table}

\begin{figure}[H]
\centering\includegraphics[scale = 0.75]{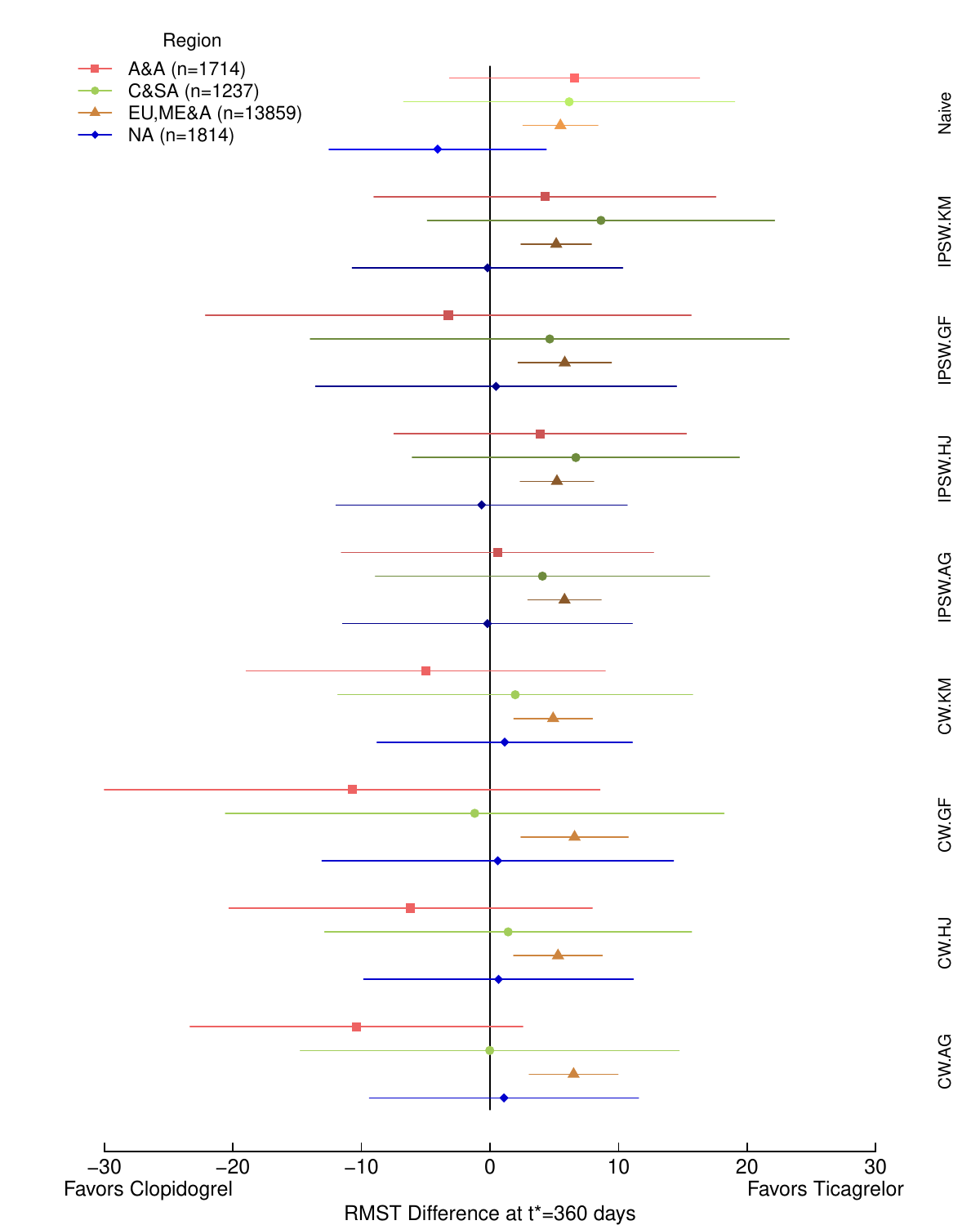}
\caption[Forest plot of average RMSTD in four-region analysis for PLATO trial.]{Forest plot of estimated average RMST differences with 95\% CIs at $t^*=360$ days from 1) A\&A: Asia and Australia, 2) C\&SA: Central and South America, 3) EU,ME\&A: Europe, Middle East and Africa, and 4) NA: North America in the four-region analysis for PLATO trial.}
\end{figure}

\newpage
\noindent Web Figure 4 presents the weighted absolute SMD of eight variables in the four-region analysis, comparing each region to the target population. The unadjusted absolute SMDs between the dominant region, Europe, Middle East, and Africa, and the target population are close to 0 for most variables except for the maintenance aspirin dosage, which is a similar finding from the two-region analysis. However, notable imbalances of the covariates are observed from the unadjusted absolute SMDs between other three regions and the target population. The CW absolute SMDs are always 0. However, the IPSW cannot well balance the distributions of maintenance aspirin dosage.

\begin{figure}[!htbp]
\centering\includegraphics[scale = 0.8]{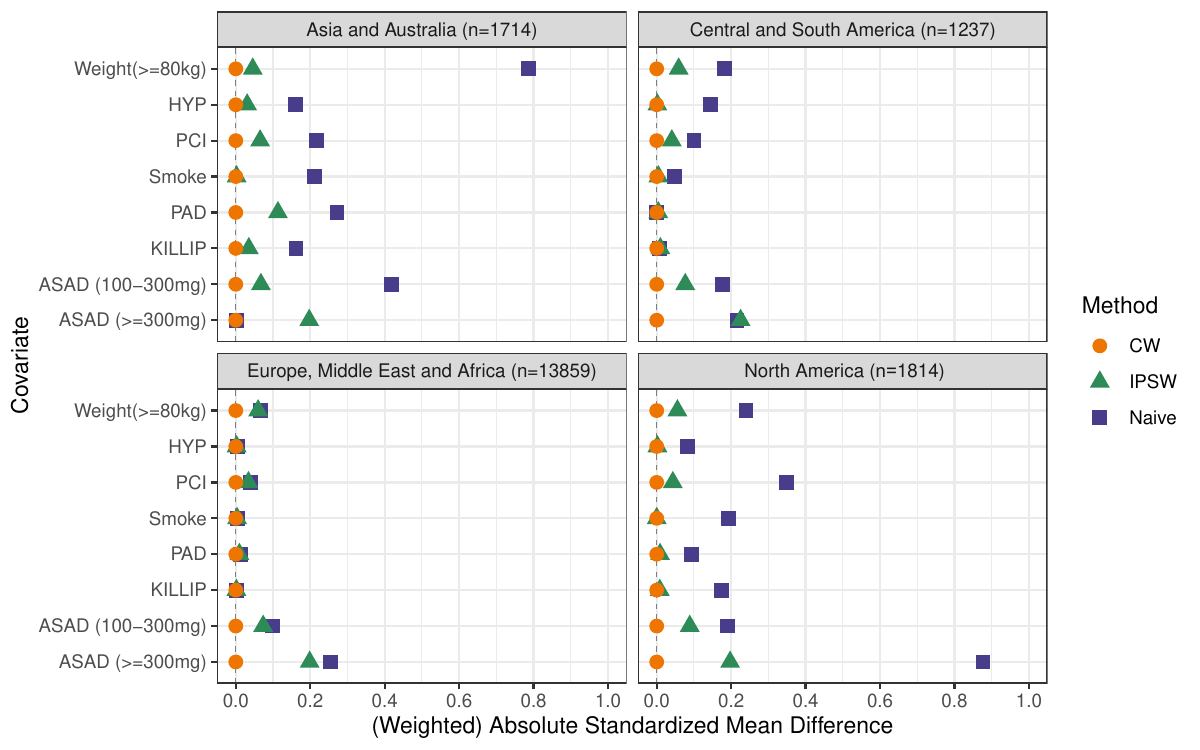}
\caption{Weighted absolute standardized mean differences of eight covariates comparing 1) Asia and Australia, 2) Central and South America, 3) Europe, Middle East and Africa, and 4) North America to the target population in the four-region analysis for PLATO trial. HYP: hypertension, PCI: percutaneous coronary intervention, PAD: peripheral arterial disease, KILLIP: Killip classification (Level I vs. Level II-IV), ASAD: aspirin dosage.}
\end{figure}

\section*{Web Appendix G: Weighted Standardized Mean Difference}
\noindent We consider the weighted absolute standardized mean difference (SMD) between variable $X$ between Region $r_1$ and $r_2$. We let $i$ index the patient and suppose the sample sizes in two regions are $n_1$ and $n_2$. We let $X_{1i}$ and $X_{2i}$ denote the individual variables of $X$ in the two regions. We let $\xi_{1i}$ and $\xi_{2i}$ denote the individual weight in the two regions.
\begin{enumerate}
    \item If $X$ is a continuous variable, the weighted absolute SMD is
    \[
    d =  \frac{|\bar{X}_1 - \bar{X}_2|}{\sqrt{s_1^2/2 + s_2^2/2}},
    \]
    where $\bar{X}_1 = \frac{\sum_{i=1}^{n_1}X_{1i}\xi_{1i}}{\sum_{i=1}^{n_1}\xi_{1i}}$ and $\bar{X}_2 = \frac{\sum_{i=1}^{n_2}X_{2i}\xi_{2i}}{\sum_{i=1}^{n_2}\xi_{2i}}$ are the weighted sample mean of $X$ in each region. $s_1^2$ and $s_2^2$ are the weighted sample variance (Galassi et al., 2002) of $X$ in each region, where 
    \[
    s_1^2 = \frac{\sum_{i=1}^{n_1}\xi_{1i}}{(\sum_{i=1}^{n_1}\xi_{1i})^2 - \sum_{i=1}^{n_1}\xi_{1i}^2}\sum_{i=1}^{n_1}\xi_{1i}(X_{1i} - \bar{X}_1)^2,
    \]
    and $s_2^2$ has the similar definition.
    \item If $X$ is a binary variable, the weighted absolute SMD is
    \vspace*{-2ex}
    \[
    d = \frac{|p_1 - p_2|}{\sqrt{p_1(1-p_1)/2 + p_2(1-p_2)/2}},
    \]
    where $p_1 = \frac{\sum_{i=1}^{n_1}X_{1i}\xi_{1i}}{\sum_{i=1}^{n_1}\xi_{1i}}$ and $p_2 = \frac{\sum_{i=1}^{n_2}X_{2i}\xi_{2i}}{\sum_{i=1}^{n_2}\xi_{2i}}$ denote the weighted prevalence of the dichotomous variable $X$ in each region.
\end{enumerate}
Note that when the weight $\xi_i = 1$ for all individuals (i.e., for the Naive unadjusted estimator), the weighted absolute SMD is equivalent to the standard absolute SMD (Austin, 2011).
